\documentclass[twocolumn,twocolappendix,tighten]{aastex631}
\usepackage{CJK}

\newcommand\J[1]{{\color{black}}}
\usepackage{soul}

\received{}
\revised{}
\accepted{}
\submitjournal{\textit{The Astrophysical Journal}}

\begin{document}

\title{Asymmetries and Circumstellar Interaction in the Type II SN~2024bch}
\correspondingauthor{Jennifer E. Andrews}
\email{jennifer.andrews@noirlab.edu}

\newcommand{\UA}{\affiliation{Steward Observatory, University of Arizona, 933 North Cherry Avenue, Tucson, AZ 85721-0065, USA}}

\newcommand{\GNL}{\affiliation{Gemini Observatory/NSF's NOIRLab, 670 N. A'ohoku Place, Hilo, HI 96720, USA}}

\newcommand{\UW}{\affiliation {DIRAC Institute, Department of Astronomy, University of Washington, 3910 15th Avenue NE, Seattle, WA 98195, USA}}

\newcommand{\keck}{\affiliation{W.M. Keck Observatory, 65-1120 Mamalahoa Highway, Kamuela, HI 96743, USA}}

\newcommand{\UCSD}{\affiliation{Department of Astronomy \& Astrophysics, University of California, San Diego, 9500 Gilman Drive, MC 0424, La Jolla, CA 92093-0424, USA}}

\newcommand{\LCO}{\affiliation{Las Cumbres Observatory, 6740 Cortona Drive, Suite 102, Goleta, CA 93117-5575, USA}}
\newcommand{\UCSB}{\affiliation{Department of Physics, University of California, Santa Barbara, CA 93106-9530, USA}}
\newcommand{\KITP}{\affiliation{Kavli Institute for Theoretical Physics, University of California, Santa Barbara, CA 93106-4030, USA}}
\newcommand{\UCD}{\affiliation{Department of Physics and Astronomy, University of California, Davis, 1 Shields Avenue, Davis, CA 95616-5270, USA}}

\newcommand{\CfA}{\affiliation{Center for Astrophysics \textbar{} Harvard \& Smithsonian, 60 Garden Street, Cambridge, MA 02138-1516, USA}}

\newcommand{\IAIFI}{\affiliation{The NSF AI Institute for Artificial Intelligence and Fundamental Interactions}}

\newcommand{\USask}{\affiliation{Department of Physics \& Engineering Physics, University of Saskatchewan, 116 Science Place, Saskatoon, SK S7N 5E2, Canada}}

\newcommand{\Rut}
{\affiliation{Department of Physics and Astronomy, Rutgers, the State University of New Jersey,136 Frelinghuysen Road, Piscataway, NJ 08854-8019, USA}}

\newcommand{\Catalyst}{\altaffiliation{LSSTC Catalyst Fellow}}

\newcommand{\JHU}{\affiliation{Department of Physics and Astronomy, The Johns Hopkins University, 3400 North Charles Street, Baltimore, MD 21218, USA}}

\newcommand{\TT}
{\affiliation{Department of Physics \& Astronomy, Texas Tech University, Lubbock, TX 79410-1051, USA}}

\newcommand{\LBI}
{\affiliation{Leibniz-Institut f{\"u}r Astrophysik Potsdam (AIP), An der Sternwarte 16, D-14482 Potsdam, Germany}}

\newcommand{\utah}
{\affiliation{Department of Physics \& Astronomy, University of Utah, Salt Lake City, UT 84112-0090, USA}}

\newcommand{\Tsinghua}
{\affiliation{Physics Department, Tsinghua University, Beijing, 100084, China}}
\author[0000-0003-0123-0062]{Jennifer E. Andrews}
\GNL
\author[0000-0002-4022-1874]{Manisha Shrestha}
\UA
\author[0000-0002-4924-444X]{K. Azalee Bostroem}
\UA\Catalyst
\author[0000-0002-7937-6371]{Yize Dong \begin{CJK*}{UTF8}{gbsn}(董一泽)\end{CJK*}}
\CfA
\author[0000-0002-0744-0047]{Jeniveve Pearson}
\UA
\author[0000-0002-9113-7162]{M. M. Fausnaugh}
\TT

\author[0000-0003-4102-380X]{David J. Sand}
\UA
\author[0000-0001-8818-0795]{S.~Valenti}
\UCD
\author[0000-0002-7352-7845]{Aravind P. Ravi}
\UCD
\author[0000-0003-2744-4755]{Emily Hoang}
\UCD
\author[0000-0002-0832-2974]{Griffin Hosseinzadeh}
\UCSD

\author[0000-0002-0551-046X]{Ilya Ilyin}
\LBI

\author[0000-0003-0549-3281]{Daryl Janzen}
\USask
\author[0000-0001-9589-3793]{M.~J. Lundquist}
\keck
\author[0000-0002-7015-3446]{Nicol\'as Meza}
\UCD

\author[0000-0001-5510-2424]{Nathan Smith}
\UA

\author[0000-0001-8738-6011]{Saurabh W.\ Jha}
\Rut

\author[0000-0002-1895-6639]{Moira Andrews}
\LCO\UCSB

\author[0000-0003-4914-5625]{Joseph Farah}
\LCO\UCSB
\author[0000-0003-0209-9246]{Estefania Padilla Gonzalez}
\LCO\UCSB
\author[0000-0003-4253-656X]{D.\ Andrew Howell}
\LCO\UCSB
\author[0000-0001-5807-7893]{Curtis McCully}
\LCO\UCSB
\author[0000-0001-9570-0584]{Megan Newsome}
\LCO\UCSB
\author[0000-0002-7472-1279]{Craig Pellegrino}
\LCO\UCSB

\author[0000-0003-0794-5982]{Giacomo Terreran}
\LCO\UCSB

\author{Patrick Wiggins}
\utah

\author[0000-0002-9454-1742]{Brian~Hsu}
\UA

\author[0000-0003-0528-202X]{Collin T. Christy}
\UA

\author[0000-0003-4537-3575]{Noah Franz}
\UA

\author[0000-0002-7334-2357]{Xiaofeng Wang}
\Tsinghua

\author[0009-0000-0314-6273]{Jialian Liu}
\Tsinghua

\author[0000-0002-9946-1477]{Liyang Chen}
\Tsinghua

\begin{abstract}
We present a comprehensive multi-epoch photometric and spectroscopic study of SN~2024bch, a nearby (19.9 Mpc) Type II supernova (SN) with prominent early high ionization emission lines. Optical spectra from \J{2.8} days after the estimated explosion reveal narrow lines of \ion{H}{1}, \ion{He}{2}, \ion{C}{4}, and \ion{N}{4} that disappear by day 6. High cadence photometry from the ground and TESS show that the SN brightened quickly and reached a peak  M$_V \sim$ $-$17.8 mag within a week of explosion, and late-time photometry suggests a $^{56}$Ni mass of 0.050 M$_{\sun}$. High-resolution spectra from day \J{7.9} and 43 trace the unshocked circumstellar medium (CSM) and indicate a wind velocity of 30--40 km s$^{-1}$, a value consistent with a red supergiant (RSG) progenitor. Comparisons between models and the early spectra suggest a pre-SN mass-loss rate of $\dot{M} \sim 10^{-3}-10^{-2}\ M_\sun\ \mathrm{yr}^{-1}$, which is too high to be explained by quiescent mass loss from RSGs, but is consistent with some recent measurements of similar SNe.  Persistent blueshifted \ion{H}{1} and [\ion{O}{1}] emission lines seen in the optical and NIR spectra could be produced by asymmetries in the SN ejecta, while the multi-component H$\alpha$ may indicate continued interaction with an asymmetric CSM well into the nebular phase.  SN~2024bch  provides another clue to the complex environments and mass-loss histories around massive stars.

\end{abstract}


\section{Introduction} 
Extensive mass-loss is an important property of massive stars, and in many cases it appears to ramp up to extreme levels  in the years to decades before core collapse \citep[see][and references therein]{2014ARA&A..52..487S}.  
While the final stages of a massive star's life may be hard to observe, some of the information is recorded in the surrounding stellar envelope and circumstellar material (CSM) and can be observed during the first few hours and days of a core collapse supernova (CCSN) explosion.  This `flash spectroscopy'
 includes fleeting narrow lines of H, He, C, and N. While this phenomenon had been well documented in several lucky cases described in the literature \citep{1985ApJ...289...52N,1994A&A...285L..13B,2000ApJ...536..239L,2007ApJ...666.1093Q,shivvers15}, within the last decade the numbers of CCSN observed early enough to detect interaction with the mass-loss from the progenitor has dramatically increased thanks to fast response by dedicated transient surveys \citep[for example:][]{GalYam2014,PTF11iqb,Yaron2017,Khazov16,2018MNRAS.476.1497B,Hosseinzadeh18,Tartaglia21,Bruch21,SN2020pni,2022ApJ...924...15J,2023ApJ...952..119B,2024ApJ...970..189J,2024ApJ...965...85A}. The recent and very nearby ($\sim$ 7 Mpc) SNe 2023ixf and 2024ggi alone have contributed many publications analyzing their early-time luminosity evolution and exquisite early time spectroscopic datasets  \citep{2023ApJ...956...46S,2024Natur.627..754L,2024Natur.627..759Z,2023ApJ...954L..42J,2023ApJ...956L...5B,2023ApJ...955L...8H,2023ApJ...954L..12T,2023SciBu..68.2548Z,2024ApJ...972L..15S,2024ApJ...972..177J,2024ApJ...970L..18Z,2024A&A...688L..28P,2024AAS...24326007D}.

The CCSNe that result from a progenitor that retains most of the hydrogen envelope prior to explosion are classified as Type II \citep[see][for detailed reviews]{2017hsn..book..239A,2017hsn..book..195G,2017suex.book.....B}. Some of these objects, particularly those with plateau light curves (SNe II-P), have been confirmed to come from red supergiant (RSG) progenitor stars via \J{archival imaging} \citep{2015PASA...32...16S,2015IAUGA..2256013V}.  When Type II SNe are observed early enough (hours to a few days after explosion) we can often observe shock interaction between the ejecta and the CSM or from the shock breakout of the SN itself \citep{Bruch21}. These fleeting emission lines show Lorentzian profiles that can have wings extending out to 1500 - 2000 km s$^{-1}$  due to electron scattering in the ionized gas (see \citealp{smith17review} for a review of SNe with dense CSM).  Early spectroscopy can be a powerful device for understanding this enhanced mass loss and its potential mechanisms prior to collapse by probing the density, temperature, and composition of the surrounding CSM.

While eruptive mass loss in luminous blue variables (LBVs)  and strong line-driven winds of  Wolf-Rayet (WR) stars are well known (see \citealp{2014ARA&A..52..487S} for a review), for RSGs there have been significant reductions in wind mass-loss prescriptions, and significant mismatch between these expected RSG wind mass-loss rates and mass-loss rates calculated from CSM interaction in SNe. For instance, the recent RSG wind mass-loss prescription from \cite{2020MNRAS.492.5994B}, which is estimated from modern high-quality IR data, reduces RSG mass-loss rates by about an order of magnitude compared to the older 1970s and 1980s literature-based canonical prescription from \citet{1988A&AS...72..259D}. Some stellar evolution models invoke even higher ``enhanced'' mass-loss rates  \citep{2012A&A...537A.146E}, but even these artificially inflated rates are still not enough to explain the $\dot{M} = 10^{-3} - 10^{-2} \ M_\sun\ \mathrm{yr}^{-1}$ estimated via flash spectroscopy observations \citep[e.g.,][]{2024ApJ...970..189J}. This discrepancy could be resolved if periods of enhanced pre-SN mass loss are considered for RSGs rather than steady, quiescent mass loss from winds. 

Causes for increased periods of mass loss could be from wave driven mass loss during late-stage nuclear burning \citep{2012MNRAS.423L..92Q,2017MNRAS.470.1642F,2021ApJ...906....3W}, turbulent convection in the core due to some dynamical instability \citep{2014ApJ...785...82S}, or even interaction with a companion and a SN progenitor's inflated envelope \citep{2014ApJ...785...82S,2013arXiv1302.5037S}, as the binary fraction for massive stars is high \citep{2012Sci...337..444S,2017ApJS..230...15M}.   Eruptive mass loss from the progenitors of normal Type II SNe have been very difficult to detect observationally \citep[for example]{2017MNRAS.467.3347K,2023ApJ...957...28D,2024ApJ...972L..15S}
suggesting if they occur they are likely some combination of dusty, faint, and quick, although see \citet{2022ApJ...924...15J} for one possible exception. 

Here we present data on SN~2024bch, a nearby and well observed Type II supernova with flash spectroscopy. 
Similar to some other SNe with flash features, SN~2024bch also shows signs of continued CSM interaction and possible asymmetries in the ejecta and/or CSM. This paper includes high cadence photometry and spectroscopy, especially in the first month after explosion, and includes high resolution echelle spectra for two epochs. In section \S\ref{sec:disc} of this paper, we outline the discovery, and the subsequent observations and data reduction is described in section \S\ref{sec:obs}.  We discuss the photometric and bolometric evolution in section \S\ref{sec:phot} and the corresponding spectroscopic evolution is described in section \S\ref{sec:spec}. A comprehensive analysis of the observations are laid out in section \S\ref{sec:analysis}, with the results summarized in section \S\ref{sec:summary}.

\begin{figure}
    \centering
    \includegraphics[width=\linewidth]{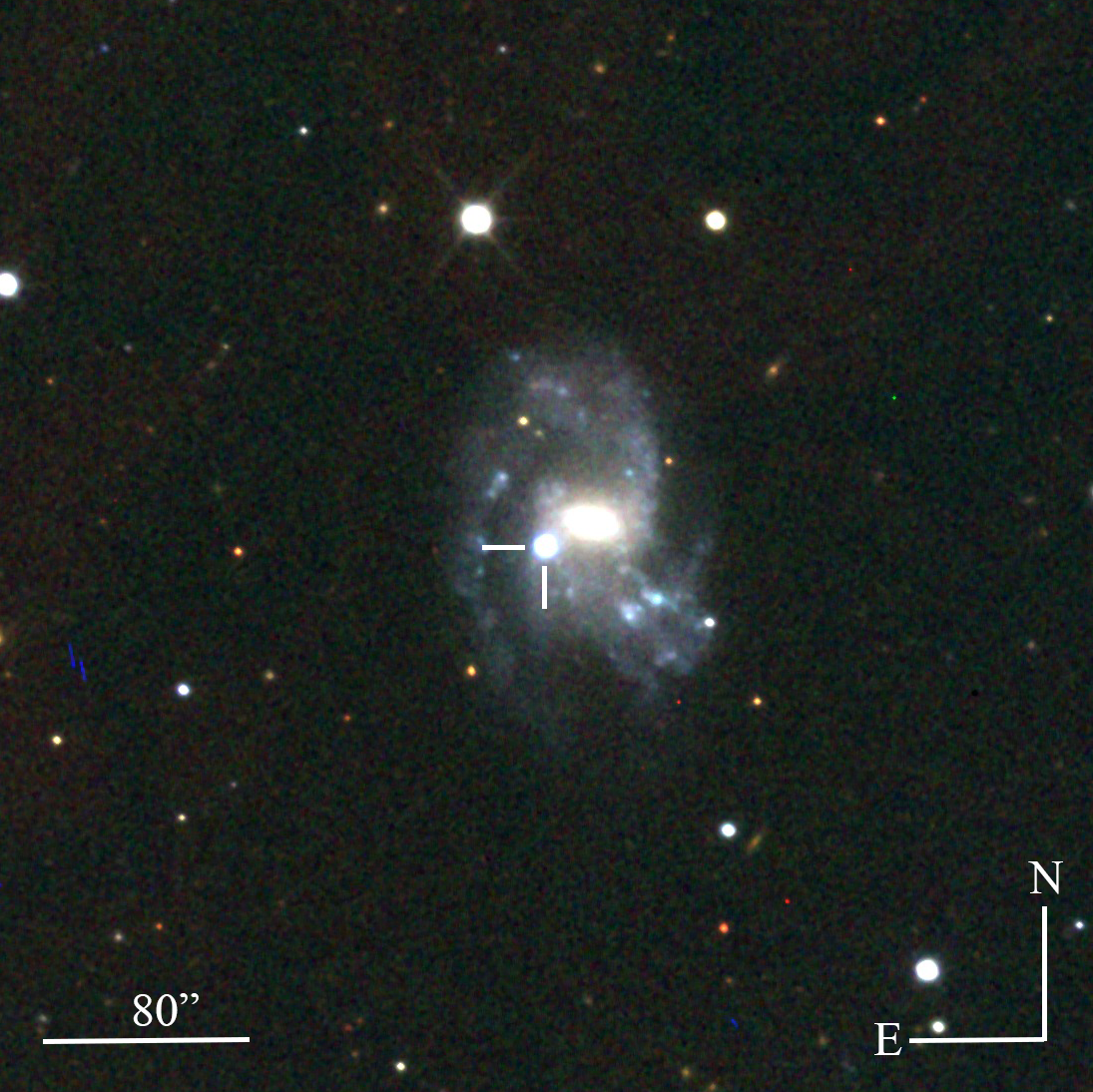}
    \caption{Composite $gri$ Las Cumbres Observatory image of SN~2024bch taken on 2024 February 1. The SN is indicated by the white cross hairs. }
    \label{fig:finder}
\end{figure}

\section{Discovery, Distance, and Reddening}
\label{sec:disc}
SN~2024bch is located in NGC~3206 (Figure \ref{fig:finder}) at J2000 coordinates $\alpha=10\textsuperscript{h}21\textsuperscript{m}49\fs742$, $\delta=+56\degr55'40\farcs53$. 
It was discovered on 2024-01-29.25 \citep[MJD 60338.25]{2024TNSTR.281....1W} and classified later that night as a possible IIn (n is for ``narrow" spectral lines), resembling the well-studied IIn SN 1998S \citep{2024TNSCR.284....1B}. While there are similarities between the two spectra, the further evolution of SN~2024bch indicated that its IIn-like features faded more quickly than SN~1998S, and that the narrow lines were likely those typically seen in objects exhibiting flash ionization. A nondetection of the object was reported by the Gravitational-wave Optical Transient Observer \citep[GOTO]{2022MNRAS.511.2405S} on 2024-01-28.04 (MJD 60337.04) with a limiting magnitude of 18.2 in their L-GOTO filter.  A day later on 2024-01-29.06 (MJD 60338.06) GOTO recorded a detection at the same magnitude of L = 18.2 $\pm$ 0.06, which was observed a few hours before the data cited in the discovery report. Without any other observations during this time to help constrain the date of explosion we will use the midway point between the GOTO limiting magnitude and the \J{ discovery, or MJD = 60337.65}, to be our explosion epoch which we will use throughout this paper. 

We also searched the Asteroid Terrestrial impact Last Alert System (ATLAS) data archive \citep{2018PASP..130f4505T,2020PASP..132h5002S}  and the Zwicky Transient Facility (ZTF) data archive \citep{2019PASP..131a8003M} at the location of SN~2024bch for any precursor outbursts using the methods presented in \citet{2023ApJ...957...28D}.  No precursor events were detected down to -11 mag in ATLAS bands for 2600 days ($\sim$7 years) or in ZTF for 2000 days ($\sim$5.5 years) prior to explosion. This is similar to what \citet{2024arXiv240915431T} found in the archival ZTF data over the past 4 years for this object. This adds to the growing number of SN with long baselines of pre-explosion photometry showing no detectable explosive progenitor history \citep[for example][]{2018MNRAS.480.1696J, 2024ApJ...972L..15S,2024MNRAS.527.5366N,2024arXiv240503747R,2024ApJ...965...93R}.  This makes the precursor detection of the only normal Type IIP SN~2020tlf \citep{2022ApJ...924...15J} even more unusual.

The heliocentric redshift of the host of SN~2024bch is z=0.003839 \citep{2011MNRAS.413.1875N}, although using the Na I D lines in our high-resolution spectra we find a value closer to z=0.003884 for SN~2024bch which we will adopt here as SN~2024bch's systemic velocity. This is a velocity offset from the host of 13 km s$^{-1}$. Using the most recent Tully-Fisher distance modulus value of $\mu$ = 31.49 $\pm$ 0.45 mag \citep{2016AJ....152...50T} gives us a distance of $19.9^{+4.5}_{-3.8}$~Mpc which we will use for this paper.
 
\begin{figure}
\includegraphics[width=\linewidth]{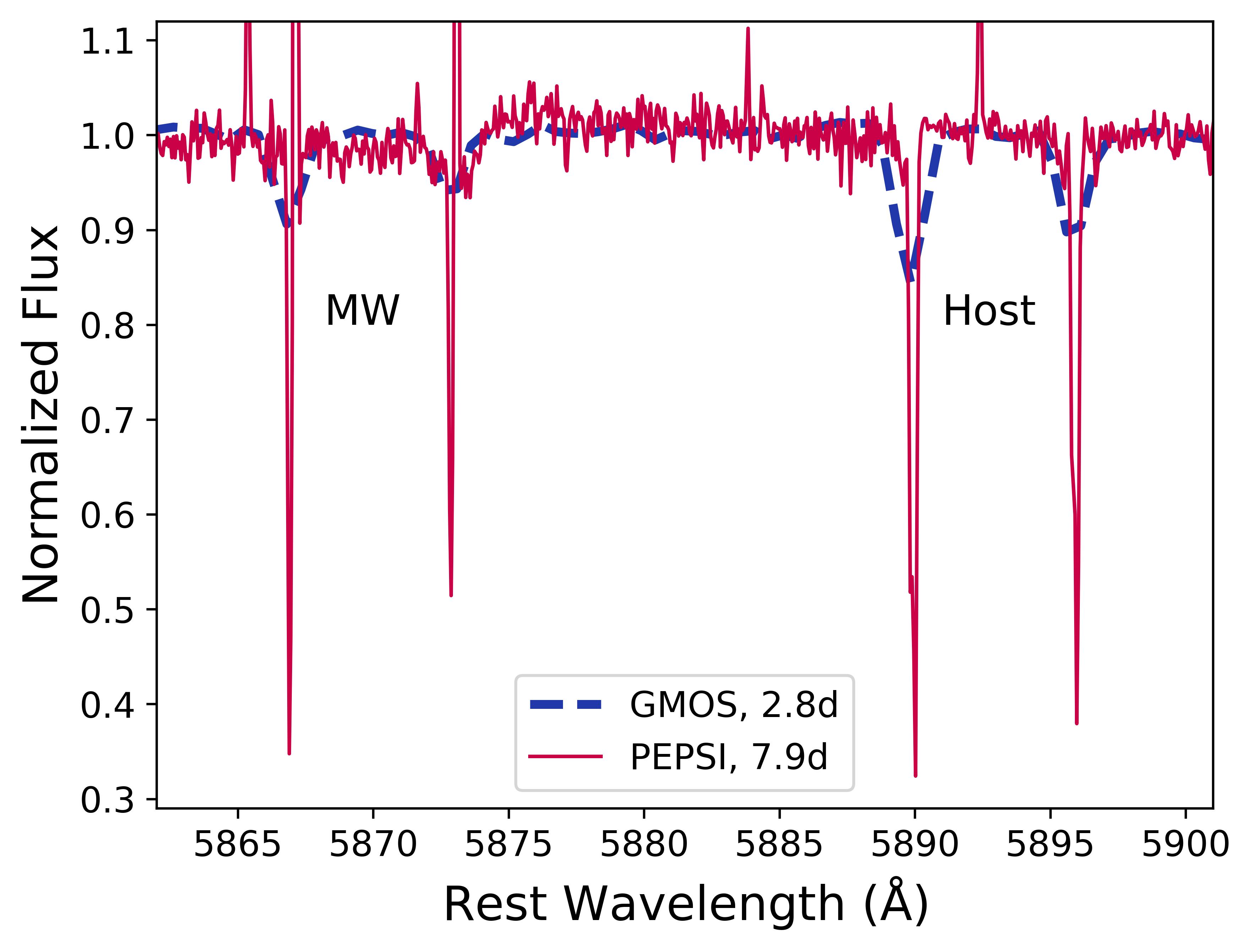}
\caption{The high-resolution day \J{7.9} PEPSI spectrum of SN~2024bch in the region around \ion{Na}{1} D. Both Milky Way and host galaxy absorption lines can be seen, as well as telluric emission lines.  For comparison, we also show the day \J{2.8} GMOS spectrum taken with the B1200 grating which shows similar, but less resolved, features. }
\label{fig:naid}
\end{figure}

In order to determine the total reddening towards SN~2024bch we use the prescription of \citet{2012MNRAS.426.1465P} and measure the equivalent width (EW) of the \ion{Na}{1} D absorption lines from the high resolution (R$\sim$50000)  spectrum taken on 2024 Feb 05 (see Section \ref{sec:specobs}). Both the Milky Way and the host galaxy \ion{Na}{1} D lines are clearly detected as shown in Figure \ref{fig:naid}, resulting in a measured $E(B-V)_{host} = 0.029 \pm 0.005 $  mag and $E(B-V)_{MW} = 0.020 \pm 0.003$ mag (applying the scaling factor of 0.86).
We do note that our Milky Way reddening measurement is slightly larger than the value of 0.013 mag derived in \citet{2011ApJ...737..103S}.
Therefore from the \ion{Na}{1} D line measurements we adopt an $E(B-V)_{tot} = 0.049 \pm 0.006$ mag, a value which we will use throughout this paper. This value is slightly larger than the one used in the recent paper on SN~2024bch from \citet{2024arXiv240915431T}, as they do not resolve any host \ion{Na}{1} D lines in their spectra and therefore use only the \citet{2011ApJ...737..103S} value. In Table \ref{tab:properties} we list all of the pertinent physical quantities for SN~2024bch.

\begin{table}
 \caption{Properties of SN~2024bch} \label{tab:properties}
 \begin{tabular}{ll}
    \hline
    Parameter & Value \\
    \hline
    R.A. (J2000) & 10$\textsuperscript{h}21\textsuperscript{m}49\fs742$  \\
    Dec. (J2000) & $+56\degr55'40\farcs53$  \\
    Last Nondetection (JD) & 2460337.54 \\ 
    Discovery (JD) & 2460338.75 \\
    Explosion Epoch (JD)\tablenotemark{a} & \J{2460338.15} \\
    Redshift ($z$)\tablenotemark{b} & 0.003884  \\
    Distance\tablenotemark{c} & $19.9^{+4.5}_{-3.8}$~Mpc\\
    Distance modulus ($\mu$) & $31.49 \pm 0.45$ mag\\
    $E(B-V)_\mathrm{MW}$\tablenotemark{d}& $0.020 \pm 0.003$ mag\\
    $E(B-V)_\mathrm{host}$\tablenotemark{d} & $0.029 \pm 0.005$ mag\\
    $E(B-V)_\mathrm{tot}\tablenotemark{d}$ & $0.049 \pm 0.006$ mag\\
    Peak Magnitude ($V_{\mathrm{max}}$) & $-$17.8 mag\\
    Rise time ($V$) &  7 days \\
    \hline
 \end{tabular}
 \tablenotetext{a}{mid point of last nondetection and \J{discovery}}
 \tablenotetext{b}{from the \ion{Na}{1}~D lines of the host galaxy}
 \tablenotetext{c}{from \citet{2016AJ....152...50T}}
 \tablenotetext{d}{from the EW of \ion{Na}{1}~D lines}
\end{table}

\section{Observations and Reductions}
\label{sec:obs}
\begin{figure*}
\includegraphics[width=\linewidth]{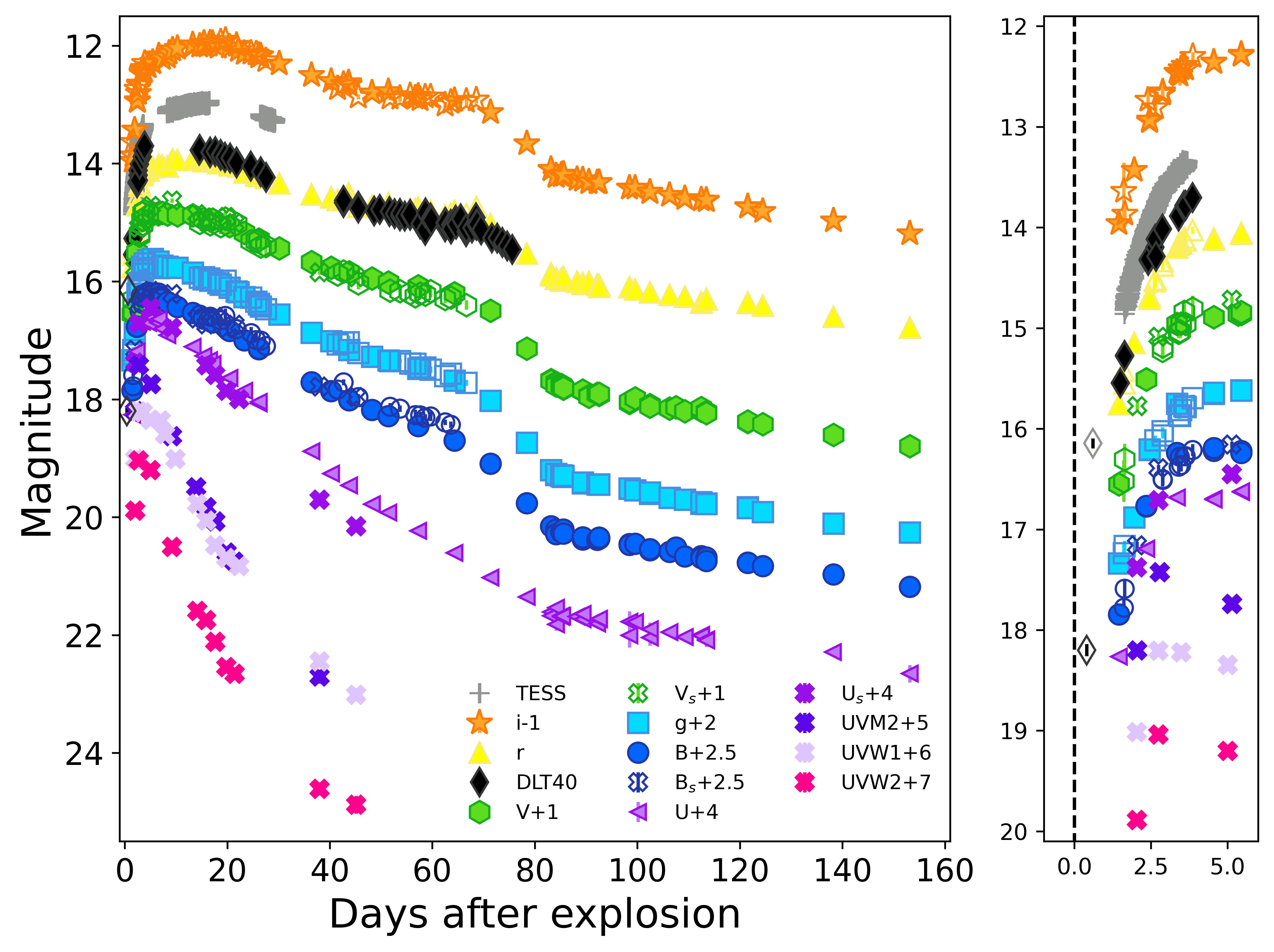}
\caption{Optical photometry of SN~2024bch, with offsets indicated in the legend. The adopted explosion epoch is MJD \J{60337.65}. Filled black diamonds and open $BVgri$ symbols are from DLT40 and the X symbols are from Swift. The open diamonds represent the reported magnitudes the first GOTO detection, and the Wiggins discovery. The right panel shows a zoom in of the first week. The dataset can be retrieved as the data behind the figure.}
\label{fig:fulllc}
\end{figure*}

\subsection{Imaging}
Photometric observations were begun soon after discovery in the $UBVgri$ bands with the 0.4-m and 1-m telescopes available through the Global Supernova Project under the Las Cumbres Observatory telescope network \citep{2013PASP..125.1031B}. PSF fitting was performed using {\tt lcogtsnpipe}, a PyRAF-based photometric reduction pipeline \citep{Valenti16}, and the $UBV$-band data were calibrated to Vega magnitudes \citep{Stetson00} using standard fields observed on the same night by the same telescope. Finally, $gri$-band data were calibrated to AB magnitudes using the Sloan Digital Sky Survey \citep[SDSS,][]{sdssdr13}. Additional high cadence photometric observations were also taken by the PROMPT-USASK telescope at Sleaford Observatory, Canada operated by the Skynet telescope network \citep{Reichart05} as part of the Distance Less Than 40 Mpc Survey (DLT40). Observations were taken in $BVgri$ and in the `Clear' filter, which was calibrated to the Sloan Digital Sky Survey $r$ band \citep[see ][for further survey and reduction details]{Tartaglia18}. Two photometric points were also collected from the Transient Name Server for epochs prior to our observing campaign as described in Section \ref{sec:disc} above.

SN~2024bch was serendipitously observed by the {\it Transiting Exoplanet Survey Satellite} \citep[{\it TESS},][]{Ricker2015}. The TESS light curve was extracted first by applying difference imaging to the TESS data, following \citet{Alard2000}, and then running forced photometry at the location of SN 2024bch in difference images. Full details of the TESS processing are given in \citet{Fausnaugh2023}. Difference imaging provides a precise differential flux of the source, but flux calibration in TESS images is uncertain by $\gtrsim$10$\%$, due TESS's large pixel that makes crowding and background estimation difficult. We instead chose to align the TESS data to the Las Cumbres $i$-band data, which has a similar pivot wavelength to the TESS filter (although the width of the filter is about half that of TESS). To calibrate the TESS data, we first averaged the TESS data over 1 hour surrounding each Las Cumbres $i$-band observation. We then solved for a rescaling factor and offset that best matches the TESS and Las Cumbres light curves in a least-squares sense. 

UV and optical images were obtained during the early portion of the light curve with the Ultraviolet/Optical telescope (UVOT; \citealp{roming05}) on board the Neils Gehrels Swift Observatory \citep[\textit{Swift}]{2004ApJ...611.1005G}. The data were downloaded from the NASA \textit{Swift} Data Archive\footnote{\url{https://heasarc.gsfc.nasa.gov/cgi-bin/W3Browse/swift.pl}}, and the images were reduced using standard software distributed with \texttt{HEAsoft}\footnote{\url{https://heasarc.gsfc.nasa.gov/docs/software/heasoft/}}. Photometry was performed for all the $uvw1$, $uvm2$, $uvw2$, $ U_\mathrm{S}$-, $ B_\mathrm{S}$-, and  $V_\mathrm{S}$-band images using a 3\farcs0 aperture at the location of SN\,2024bch. Since no pre-explosion template imaging was available, the contribution from the host galaxy, which was insignificant, has not been subtracted. All of the lightcurves are shown in Figure \ref{fig:fulllc}.

\subsection{Spectroscopy}
\label{sec:specobs}
Multiple epochs of optical spectroscopy were obtained with the FLOYDS spectrograph \citep{2013PASP..125.1031B}
on the Las Cumbres Observatory's 2m Faulkes Telescopes North (FTN) as part of the Global Supernova Project collaboration. One-dimensional spectra were extracted, reduced, and calibrated following standard procedures using the FLOYDS pipeline \citep{Valenti14}. Additionally, multiple epochs were obtained with the Gemini Multi-Object Spectrograph (GMOS; \citealp{hook04,gimeno16}) on the 8.1\,m Gemini North Telescope using the B480 and B1200 gratings. Data were reduced using the Data Reduction for Astronomy from Gemini Observatory North and South reduction package \citep[{\tt DRAGONS},][]{2023RNAAS...7..214L}, using the recipe for GMOS long-slit reductions. This includes bias correction, flatfielding, wavelength calibration, and flux calibration. Data was also taken on the Boller $\&$ Chivens (B$\&$C) Spectrograph on the Bok 2.3 m Telescope, and were reduced using standard IRAF techniques. We also obtained spectra
with the Kast spectrograph on the 3m Shane telescope at Lick Observatory. Kast spectra were reduced using standard IRAF/Pyraf \citep{2012ascl.soft07011S} and Python routines
for bias/overscan subtractions and flat-fielding. Two epochs of high resolution (R$\sim$50000) echelle spectra were taken with the Potsdam Echelle Polarimetric and Spectroscopic Instrument \citep[PEPSI,][]{2018SPIE10702E..12S} mounted on the Large Binocular Telescope (LBT) located on Mt.\,Graham, AZ. The data were reduced using the Spectroscopic Data Systems pipeline \citep{2018SPIE10702E..12S}, in the same manner as described by \citet{2023ApJ...956...46S}. One spectrum was taken with the Low-Resolution Imaging Spectrometer \citep[LRIS;][]{Oke1995} on the 10~m Keck~I telescope and reduced using the LPipe package \citep{2019PASP..131h4503P}. Three spectra were obtained with the  Beijing Faint Object Spectrograph and Camera (BFOSC) mounted on the 2.16~m telescope at Xinglong observatory, China. Finally, two spectra were also downloaded from the Transient Name Server to add to our analysis including the classification spectrum from \citet{2024TNSCR.284....1B} and an additional spectrum from day 3.4 from the CAFOS instrument mounted on the CA-2.2m telescope in Spain. A log of the optical spectroscopic observations can be found in Table \ref{tab:optspec}.

Four epochs of NIR spectroscopy were obtained with SpeX \citep{2003PASP..115..362R} on the NASA Infrared Telescope Facility (IRTF) and the MMT and Magellan Infrared Spectrograph \citep[MMIRS;][]{mmirs} on the 6.5~m MMT located on Mt.\,Hopkins in Arizona. SN\,2024bch was observed with IRTF on two seperate nights using an ABBA dithering pattern. The associated flat-field and comparison arc-lamp observations were taken right after the science observation cycles. The IRTF SpeX data were reduced with Spextool \citep{Cushing04} and the output was telluric corrected using a standard A0V star observed at similar airmass adjacent to the science target, following the prescription in \cite{Vacca03}. The two epochs of MMT data were manually reduced using the MMIRS pipeline \citep{mmirspipe}, then the 1-D spectral outputs were telluric and absolute flux corrected following the method described in \citet{Vacca03} with the \verb|XTELLCOR_GENERAL| tool \citep[part of the Spextool package]{Cushing04} using a standard AV0 star observed at similar airmass and time. The summary of NIR spectra are listed in Table \ref{tab:nirspec}.

\section{Photometric Evolution}
\label{sec:phot}
\begin{figure}
    \centering
    \includegraphics[width=\linewidth]{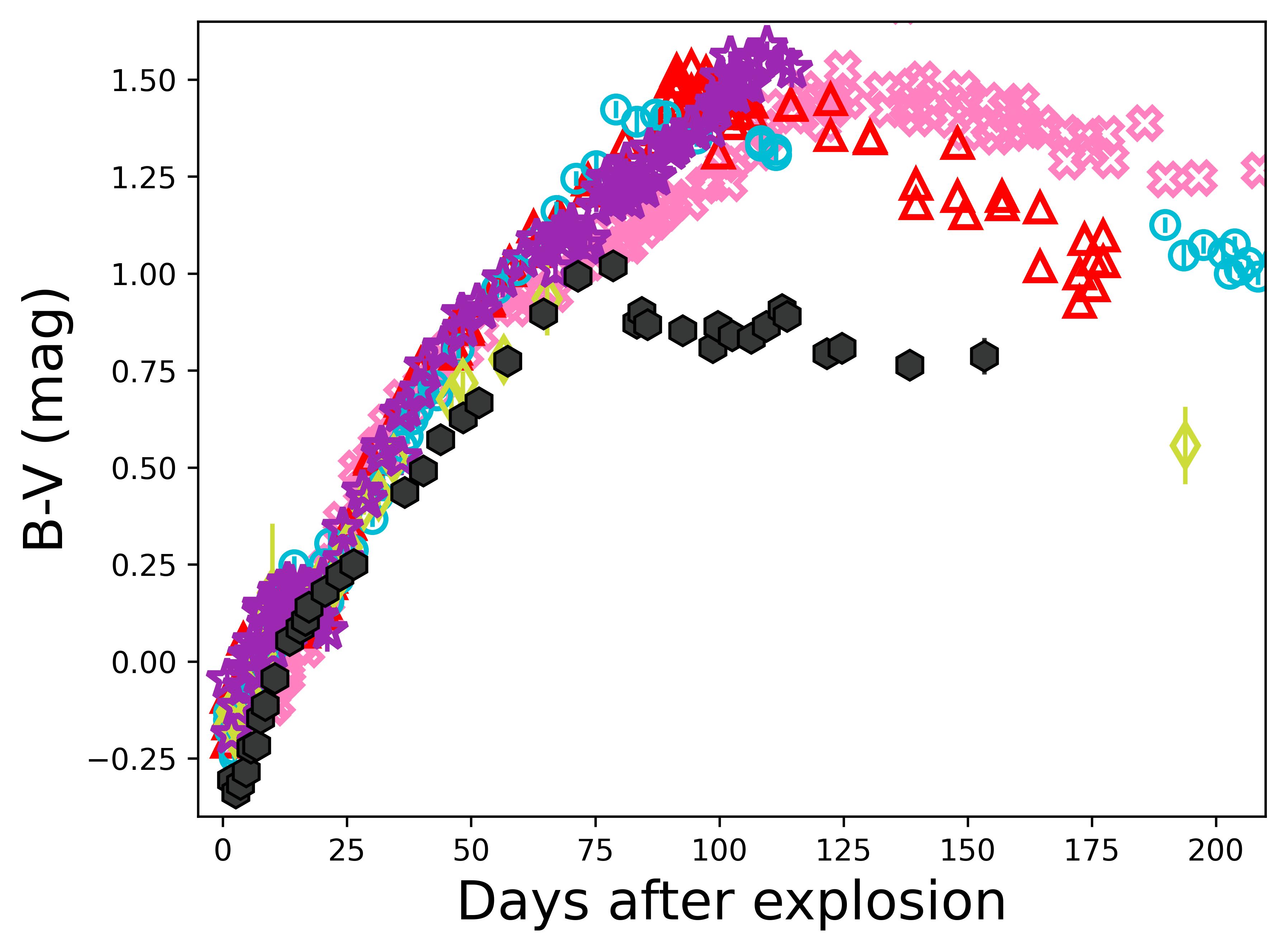}
     \includegraphics[width=\linewidth]{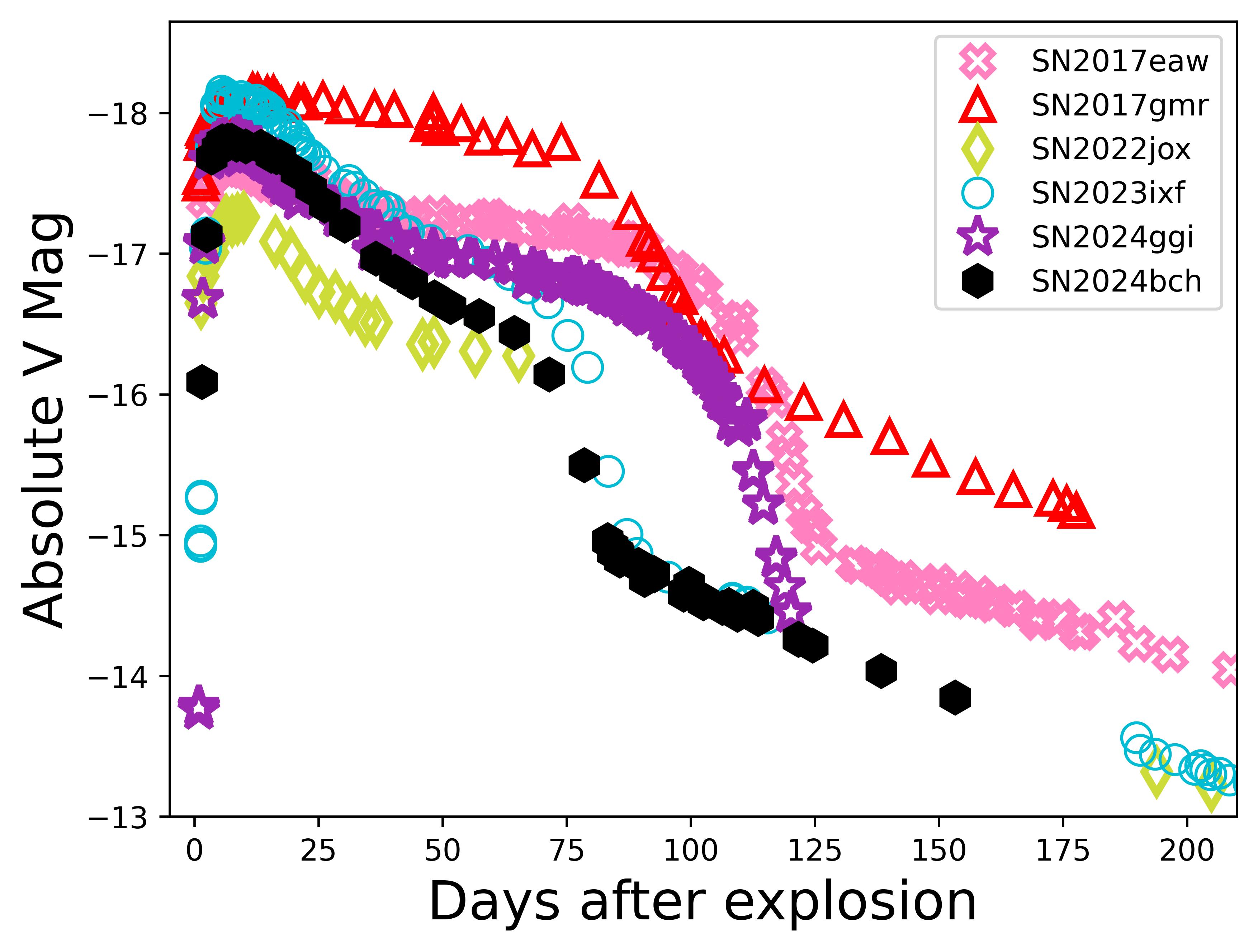}
    \caption{Top: $B-V$ color evolution of SN~2024bch compared with other Type II SNe \J{selected as described in the text}. Bottom: Absolute V-band photometry of the same objects above.  Data are from  \citet[SN~2017eaw]{2019ApJ...876...19S}, \citet[SN~2017gmr]{Andrews2019}, \citet[SN~2022jox]{2024ApJ...965...85A}, \citet[SN~2023ixf]{2024arXiv240807874H}, and \citet[SN~2024ggi]{2024ApJ...972L..15S}.  Additional SN~2024ggi data obtained from Las Cumbres. All lightcurves have been corrected for reddening and distances from values within. Legend is the same for both figures.}
    \label{fig:colorcompare}
\end{figure}

\begin{figure}
    \centering
    \includegraphics[width=\linewidth]{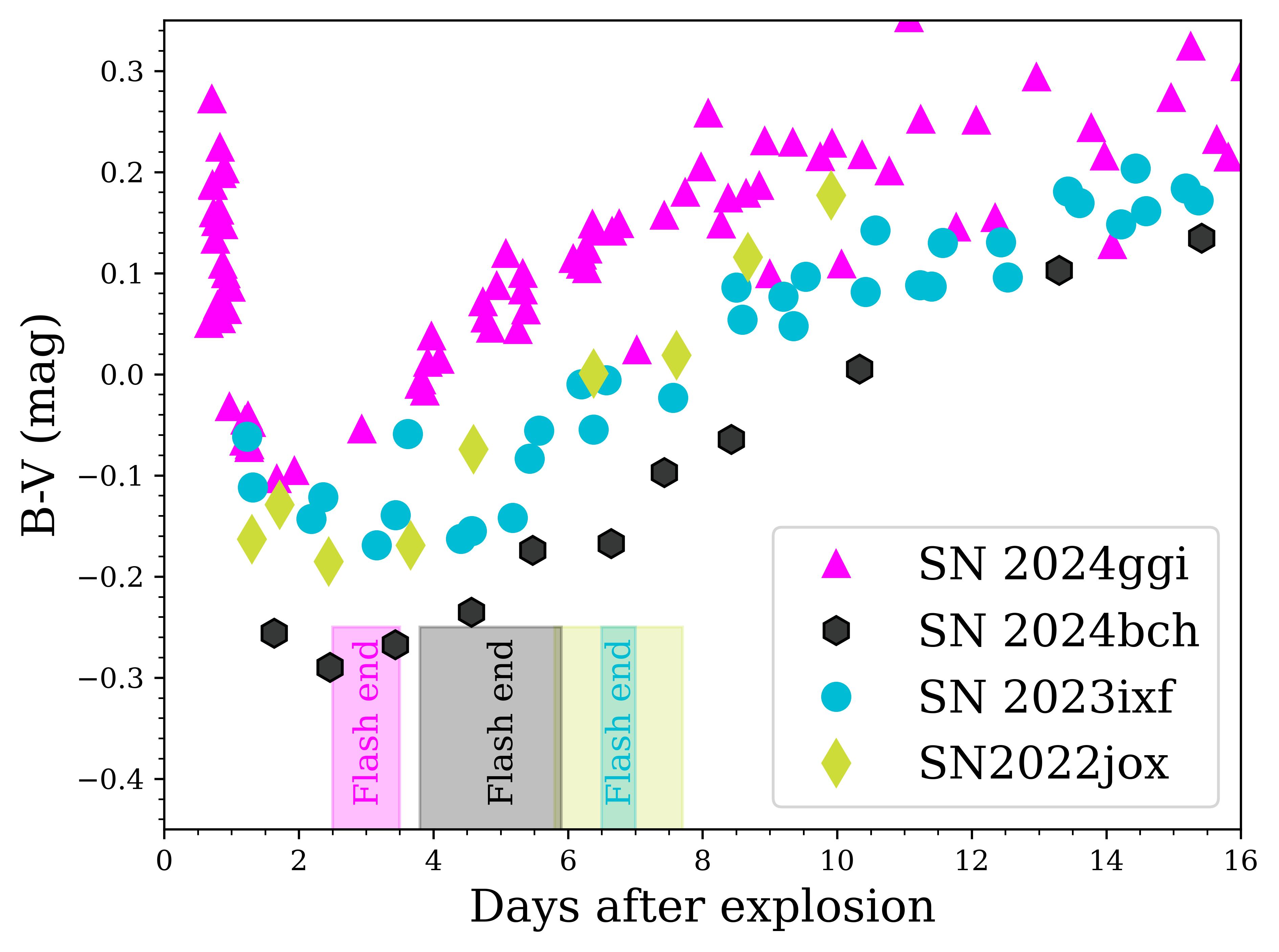}
    \caption{Extinction corrected $B-V$ plot adapted from \citet{2024ApJ...972L..15S} with the addition of SN~2024bch and SN~2022jox. The last epoch where spectra have flash features are indicated by the shaded regions corresponding to symbol color. As in their Figure 8, we have made the position and width of the shaded region to correspond to the time between the last spectrum showing a flash feature and the first spectrum with no flash features.  }
    \label{fig:colorflash}
\end{figure}
\begin{figure}
 \centering
\includegraphics[width=3.5in]{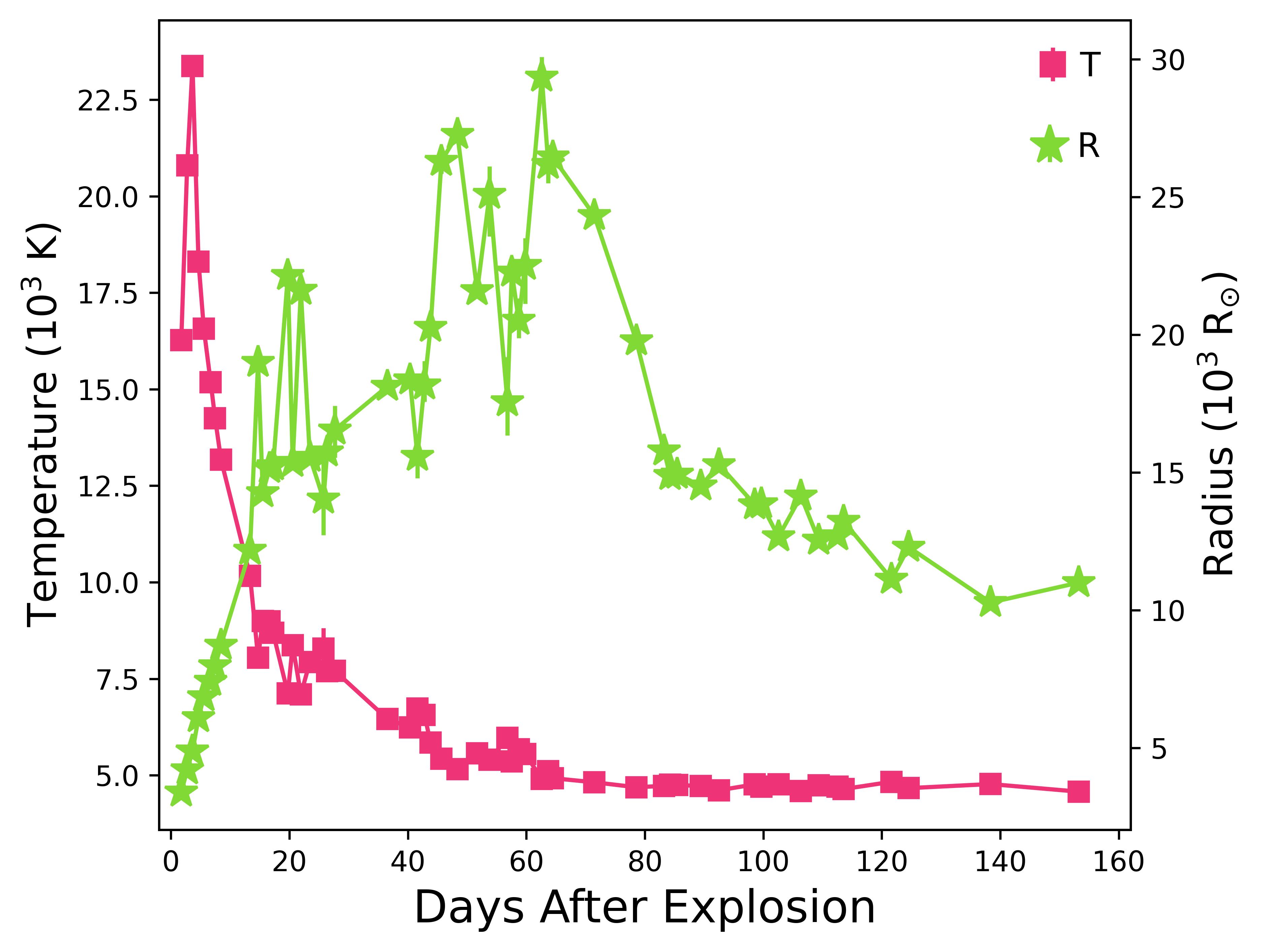}
\includegraphics[width=3.2in]{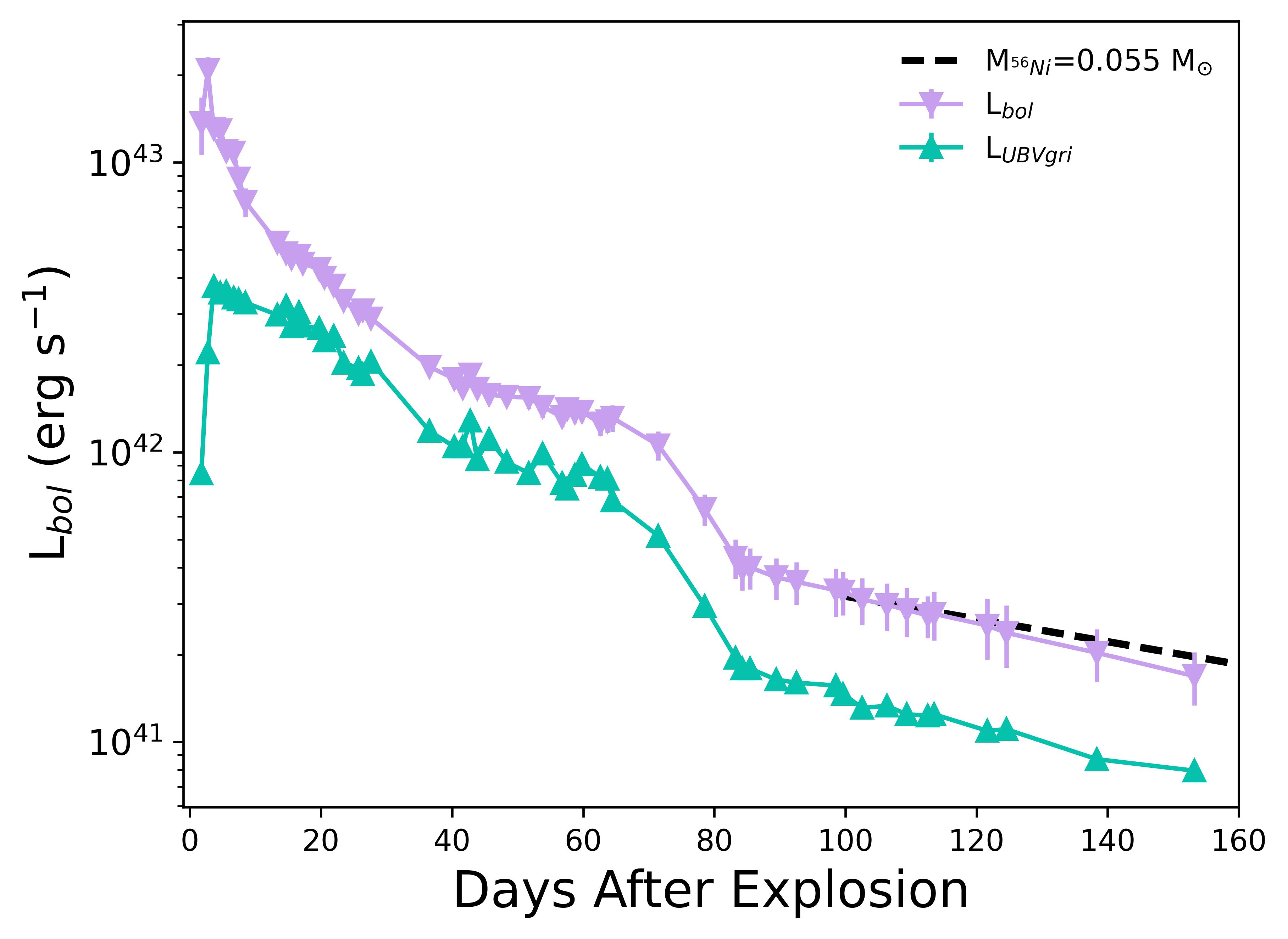}
\caption{ Blackbody temperature, radius (top), and luminosity (bottom) evolution of SN~2024bch derived from the UV to optical photometry.  All data have been dereddened by our assumed $E(B-V)_{tot}$ = 0.049 mag. Temperature and radius are derived from fitting the Planck function to the photometry using an MCMC routine.  Both the derived bolometric (purple) and psuedo-bolometric (cyan) luminosity are shown.}
\label{fig:RTTot}
\end{figure}

\subsection{Optical and UV Lightcurves}
The full optical and UV lightcurves of SN~2024bch are shown in Figure \ref{fig:fulllc}, and the V-band lightcurve in absolute magnitude is shown in the bottom of Figure \ref{fig:colorcompare}. Our estimated explosion epoch of MJD = \J{60337.65} is indicated by the dashed line in the right panel of Figure \ref{fig:fulllc}. We were able to observe fairly continuously from $\sim$ 1.5 days out to $\sim$ 155 days before the SN became Sun constrained. From our data it appears that the V band reaches a maximum $M_{V} = -17.8$ mag at roughly 7 days post explosion, while the  $r$ and $i$-band takes much longer, roughly 15 to 17 days to reach the maximum. This could potentially be due to CSM interaction that will be discussed in detail below, but is still within somewhat normal values for some Type II SNe \citep{2014ApJ...786...67A,2015MNRAS.451.2212G}. The $UBg$ observations also show a peak at roughly 5 days, once again on the shorter end of the distribution, but within the normal range. We also catch a rise in the Swift data between our first and second epoch indicating maximum was reached somewhere near day 3. Unfortunately, the cadence is not very high for our Swift observations and our first epoch was at \J{2.2} days after explosion so we cannot constrain the UV behavior more than this.

As can be seen in Figure \ref{fig:fulllc}, the plateau length persists only until $\sim$ 70 days post-explosion, which is a bit on the shorter side for Type II-P supernovae \citep{2014ApJ...786...67A,Valenti16}. The fall from the plateau lasts about 15-20 days (the observational cadence was quite low during this time), and when the radioactive tail phase begins $M_{V} = -14.8$ mag, or about 3 magnitudes fainter than at peak (bottom panel Figure \ref{fig:colorcompare}). 

\J{In Figure \ref{fig:colorcompare} we compare the $B-V$ color and $M_{V}$ evolution of SN~2024bch to other nearby Type II SNe.  We chose objects specifically with high observational cadence and with early enough observations to determine whether narrow lines were present or not. Additionally, we selected SN that are at a similar peak brightness to SN~2024bch.} Overall the shape and brightness of the V-band lightcurve is quite similar to the recent and well-studied Type II SN~2023ixf, both in plateau duration and shape and in the post-plateau brightness \citep{2024arXiv240807874H}. The lightcurves of SN~2024ggi and SN~2022jox, other recent CCSNe showing flash features, also have very similar properties at least until about 50 days when the lightcurve of SN~2024ggi shows a much more pronounced and prolonged plateau. SN~2017gmr and SN~2017eaw, and the second half of the plateau phase for SN~2024ggi, show the more classical plateau shape and length. Interestingly, both SN~2017gmr and SN~2017eaw were discovered quite early and have spectroscopy within the first few days and do not show flash features other than possibly fleeting narrow H$\alpha$ and broad He bumps \citep{Andrews2019, 2019MNRAS.485.1990R}, and SN~2024ggi only showed flash features until day 3-4 \citep{2024ApJ...972L..15S,2024ApJ...970L..18Z,2024ApJ...972..177J}.

\subsection{Color Evolution}

In the top panel of Figure \ref{fig:colorcompare} we show the $B-V$ color evolution of SN~2024bch compared with other well studied Type II SNe. \J{A zoom in of the first few days for those comparison SNe showing flash features are shown in Figure \ref{fig:colorflash}}. As seen from the figures, we are able to catch a blueward color evolution over the first three days, reaching $B-V = -0.3$ mag, and then a slow evolution to the red until reaching a value of $B-V$ = 1.0 mag around day 75, when the fall from plateau begins.  From this point onward there is a quick drop to $B-V$ = 0.9 mag and the color slowly trends blueward with a $B-V$ = 0.8 mag around day 150 when the SN becomes Sun constrained. 

As SNe are being discovered closer and closer to the explosion epoch, we have seen similar early time color behavior in other Type II SNe \J{with flash features} as shown in Figure \ref{fig:colorflash}.  As mentioned in \citet{2024ApJ...972L..15S}, SN~2024ggi and SN~2023ixf both show an evolution to a $B-V$ minimum within days of explosion, just like SN~2024bch and SN~2022jox. When we compare this color evolution with the timescales of flash duration there does seem to be a loose correlation; the faster the transition to redward color evolution the earlier the flash features will disappear. 

\subsection{Bolometric Lightcurve and $^{56}$Ni Mass} 

The bolometric and pseudobolometric lightcurves of SN~2024bch calculated from the Light Curve Fitting Package \citep{2023zndo...7872772H} are shown in the bottom panel of  Figure \ref{fig:RTTot}. To construct the bolometric lightcurve, an MCMC fitting routine fit a blackbody spectrum to each epoch of observed SED\footnote{While the blackbody fit is not strictly valid at late times when spectral lines have emerged, we compared our pseudobolometric light curves derived from this method and from direct integration of the observed optical SED and found the difference to be negligible.}.  Assuming the reddening and distance listed in Table \ref{tab:properties}, we estimate a maximum bolometric luminosity of L$_{bol}$ = 2.1 $\times$ 10$^{43}$ erg s$^{-1}$ roughly 3 days after explosion. The pseudobolometric lightcurve is made from the best-fit blackbody integrated from $U$ to $i$ for each epoch and peaks at a value of  L$_{bol}$ = 3.6 $\times$ 10$^{42}$ erg s$^{-1}$.

The late-time bolometric luminosity appears to decline at a rate of 0.014 mag d$^{-1}$, a value higher than the canonical rate of fully-trapped $^{56}$Co decay of 0.0098 mag d$^{-1}$. This is shown as the dashed line in the bottom of Figure \ref{fig:RTTot} where we show the decay of a mass of $^{56}$Ni = 0.055 M$_{\sun}$ (a value we derive below) for comparison.  The slight deviation and quicker decline of SN~2024bch may indicate that the gamma rays were not fully trapped \citep{2014ApJ...786...67A,2022A&A...660A..42M}, that there is potentially dust formation occurring, or that there is CSM interaction which is contributing to the overall luminosity and not only increasing the luminosity but fading at a higher rate than radioactive decay alone. While we will explore these in more detail later, average values for $^{56}$Ni masses in normal Type II SNe have been found to be $\sim$ 0.04 M$_{\sun}$ \citep{Valenti16,2017ApJ...841..127M,2019A&A...628A...7A, 2021MNRAS.505.1742R}. The slightly larger estimated $^{56}$Ni mass combined with the shape of the H$\alpha$ emission (described in section \ref{sec:halpha}) during the same time may indicate that CSM interaction may be the reason for this discrepancy.

For our $^{56}$Ni mass estimates we use the prescriptions described in \citet{Hamuy2003} and \citet{Jerkstrand2012} both of which employ the radioactive tail of the bolometric lightcurve. We estimate the value at various epochs and find that on day 103 we get $^{56}$Ni masses of 0.058 M$_{\sun}$ $\pm$ 0.002 and 0.055 M$_{\sun}$ $\pm$ 0.002 respectively.  These values fall to $^{56}$Ni masses of 0.050 M$_{\sun}$ $\pm$ 0.003 and 0.047 M$_{\sun}$ $\pm$ 0.003 on day 153. We, therefore, estimate the $^{56}$Ni mass to be 0.05 M$_{\sun}$ with the caveat that extra heating from late CSM interaction can inflate this value and that incomplete gamma ray trapping is not taken into account. 

We also calculate the temperature and photospheric radius using the Light Curve Fitting Package shown in the top panel of Figure \ref{fig:RTTot}. The first few epochs reveal a rise in temperature T$_{BB}$ from 15.0 kK on \J{1.6} days to a max of 23.4 kK on day \J{3.4}. \J{When compared to a small sample of other SNe with flash features}, this is significantly cooler than the maximum temperatures of SN~2023ixf of 34.4 kK \citep{2024Natur.627..759Z} and 30.7 kK for SN~2024ggi \citep{2024ApJ...972L..15S} but consistent with the 25 kK max for SN~2022jox \citep{2024ApJ...965...85A}. Note that the cadence of Swift observations were different for all four SNe, which could contribute to some of the differences as the bulk of the luminosity is in the UV at this point. There is then a general decline of T$_{BB}$ from max over the next 60--80 days until reaching a value of 4.75 kK as it moves to the radioactive tail. The photospheric radius, R$_{BB}$, rises to a maximum of roughly 2.9 $\times$ 10$^{4}$ R$_{\sun}$ during the first 60 or so days as the SN is on the plateau.

\section{Spectroscopic Evolution}
\label{sec:spec}
\subsection{Optical Spectra}
\begin{figure*}
\includegraphics[width=\linewidth]{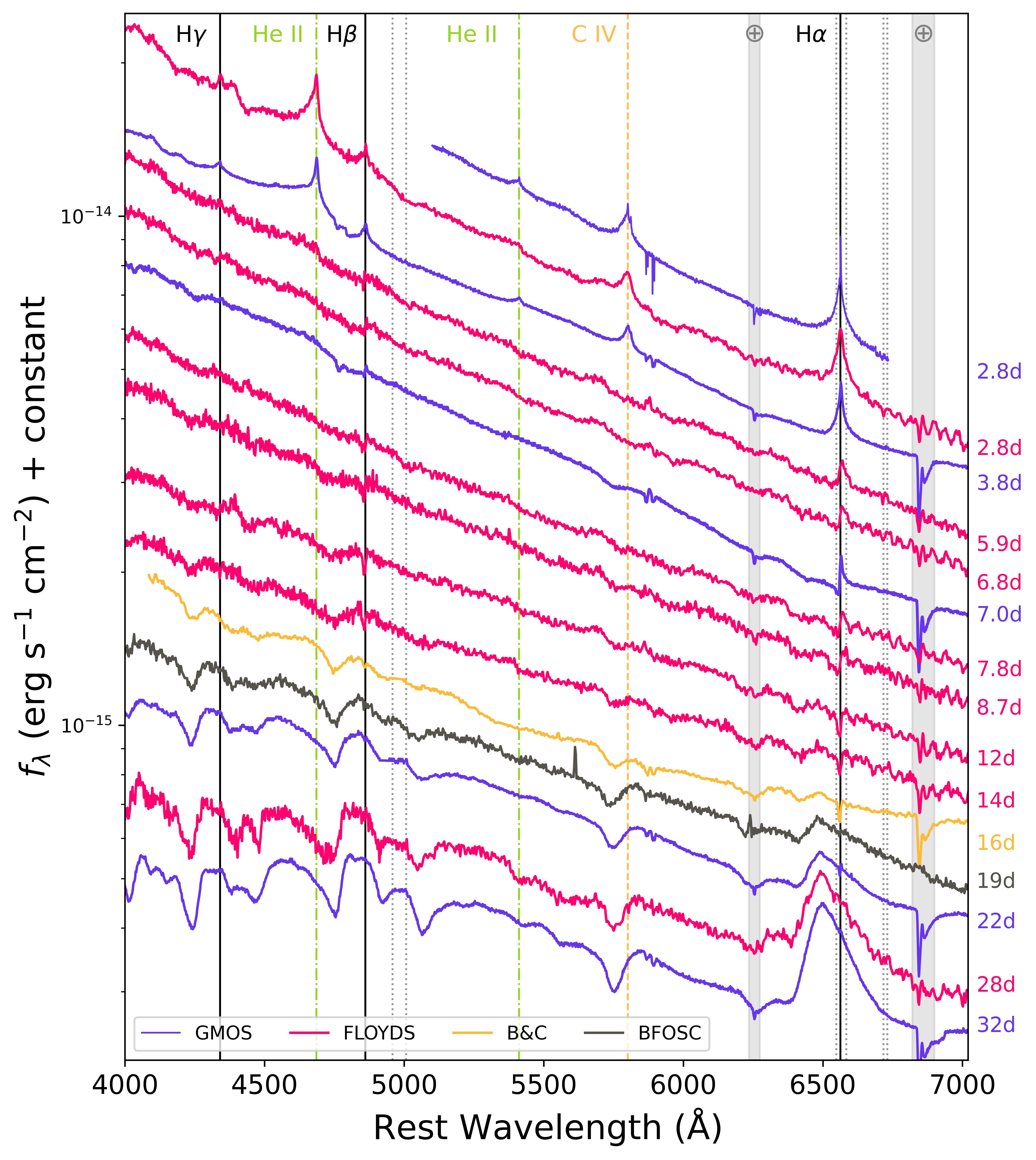}
\caption{Optical spectra of SN~2024bch for the first month after explosion.  Each telescope+instrument pair is notated by a different color. Notable lines are identified, and H{\sc~ii} region lines are marked with dotted gray lines. The dates are with respect to our assumed explosion epoch of MJD \J{60337.65}. }
\label{fig:earlyspec}
\end{figure*}

\begin{figure*}
\includegraphics[width=\linewidth]{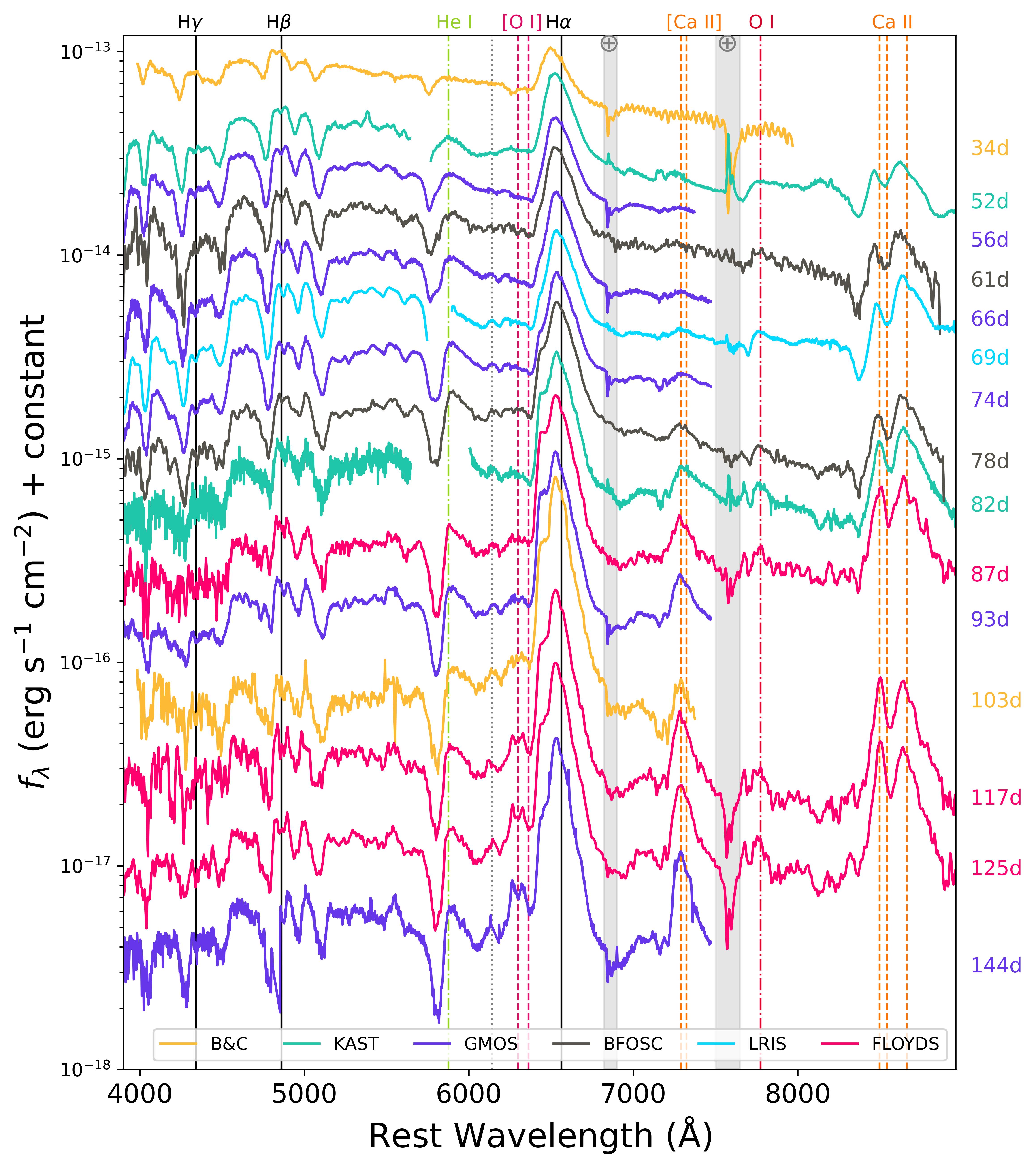}
\caption{Same as Figure \ref{fig:earlyspec} but for days 34--144 d.  Each telescope+instrument pair is notated by a different color. Notable lines are identified and the dates are with respect to our assumed explosion epoch of MJD \J{60337.65}. }
\label{fig:latespec}
\end{figure*}

As summarized in Table \ref{tab:optspec} and shown in Figures \ref{fig:earlyspec} and \ref{fig:latespec}, we obtained optical spectra from \J{2.8} days after explosion until day 147 when SN~2024bch was lost behind the Sun.  On day \J{2.8} we have both low- and medium-resolution data, which show \ion{H}{1} emission, along with the high-ionization lines of \ion{He}{2} $\lambda$4686 and \ion{C}{4} $\lambda\lambda$5801,5811. 
Note in our higher-resolution GMOS B1200 data we are able to easily resolve the separate \ion{C}{4} lines as well as \ion{He}{2} $\lambda$6560 which is often blended with H$\alpha$. Absent are the lines of \ion{N}{3} $\lambda\lambda$4634,4641, \ion{C}{3} $\lambda\lambda$4648,4650 and \ion{N}{5} $\lambda\lambda$7109,7123 which has been seen in the early spectra of other Type II SNe. This does not discount their presence at earlier epochs, but by \J{2.8} days if they had existed they have faded away. In the day 1.40 and 1.42 spectra presented in \citet{2024arXiv240915431T} there is a feature blueward of \ion{He}{2} $\lambda$4686 that has been identified as \ion{N}{3}; this is likely correct but the resolution makes it difficult to definitively confirm.  These high ionization lines all disappear after our \J{3.8} day spectrum and before our \J{5.9} day spectrum. The Balmer emission lines persist for a few more days after the other high ionization lines are gone. Combining an ejecta velocity of 7000 km s$^{-1}$ measured from H$\beta$ and \ion{Fe}{2} $\lambda$5169 from our day 16 and 22 spectra and the range of dates when the disappearance of high ionization lines occurred we can estimate an approximate CSM radius of R$_{CSM}$ = 2.4 -- 3.6 $\times$ 10$^{14}$ cm.   Assuming a progenitor wind velocity of 35 km s$^{-1}$ that we measure from the high resolution spectrum (and discussed in detail below), this quiescent mass loss would have occurred  2.4--3.6 years prior (much earlier than the 9 years derived in \citet{2024arXiv240915431T}). It is interesting to note though that no obvious outburst was seen during this time period in the ATLAS and ZTF precursor imaging.

During the second week, the only distinguishable features are narrow absorption lines in H$\alpha$ and H$\beta$.  In our lower resolution FLOYDS spectrum on day 12 and slightly higher resolution Bok spectrum on day 16, the absorption lines have FWHM = 400--700 km s$^{-1}$ with centers blueshifted from zero velocity.
Around three weeks after explosion the photosphere begins to cool and broad emission lines from the fast moving ejecta start to appear, including a prominent H$\alpha$ with rather weak P-Cygni absorption with a minimum at $-7500$ km s$^{-1}$.  In the day 22 spectrum, the narrow H$\alpha$ absorption has become a narrow P-Cygni profile with the reappearance of a narrow emission feature. There also looks to be a high velocity absorption line at $-14000$ km s$^{-1}$, but this is more likely a telluric line as described in \citet{2000AJ....120.1499M}. In section \ref{sec:halpha} below we discuss the H$\alpha$ evolution in more detail. 

After about a month, the \ion{Ca}{2} IR triplet emerges, along with \ion{He}{1} and various Fe lines, including \ion{Fe}{2} $\lambda\lambda$4924,5018, and $\lambda$5169. After the fall from the plateau, starting around day 80, the [\ion{Ca}{2}] emission lines strengthen substantially,
and \ion{Ba}{2} $\lambda$6142, [\ion{Sc}{2}], and \ion{He}{1} $\lambda$7065 start to appear in the nebular spectra. The [\ion{O}{1}] and \ion{O}{1} $\lambda$7774 lines stay fairly suppressed until after 100 days, and when the [\ion{O}{1}] $\lambda$$\lambda$6300,6363 doublet does become pronounced, the peaks are blueshifted at roughly the same velocity of H$\alpha$ of $-1500$ km s$^{-1}$. The implications of this will be explored more in section \ref{sec:asym}.

\subsection{High-resolution spectra}

\begin{figure}
\includegraphics[width=\linewidth]{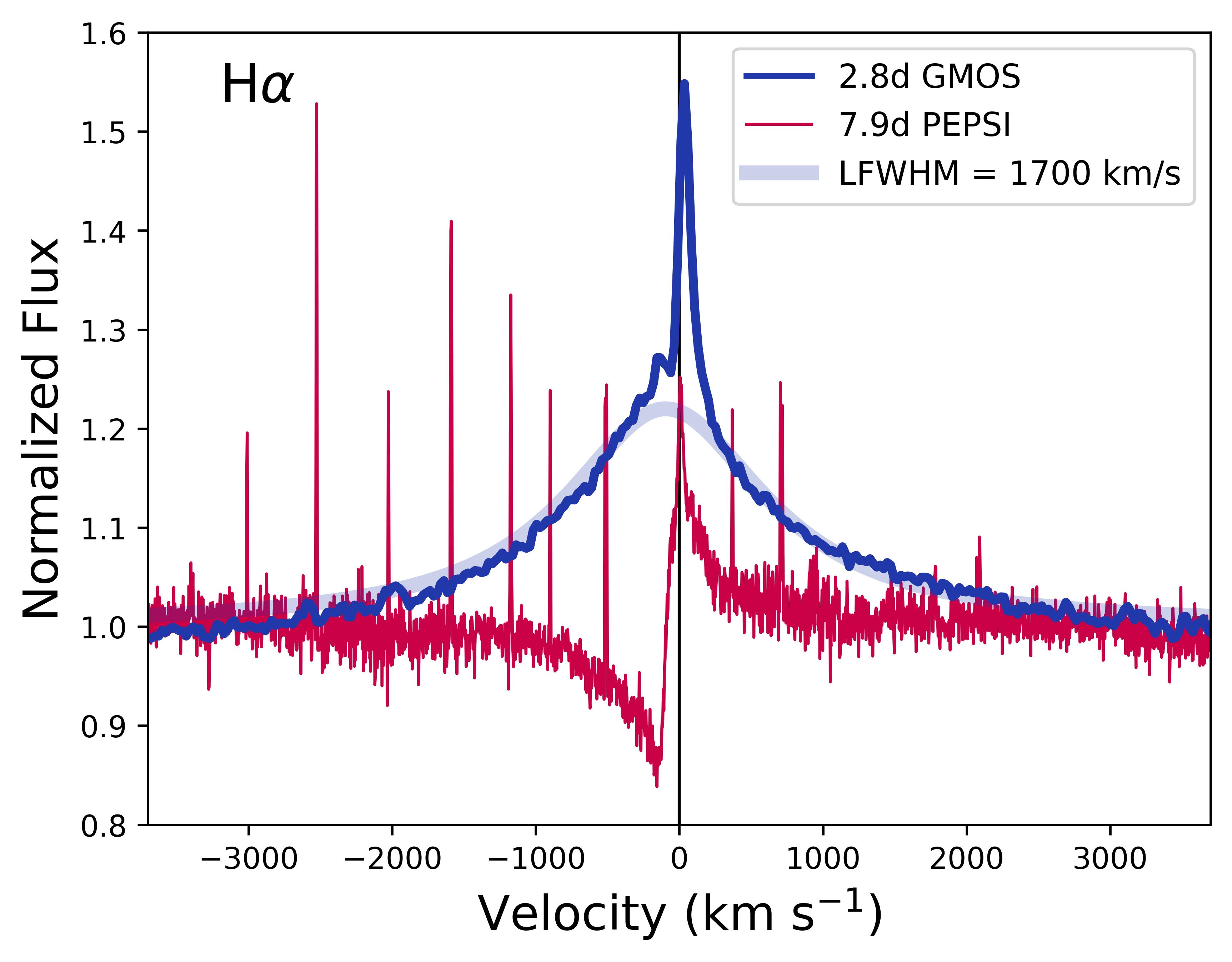}
\includegraphics[width=\linewidth]{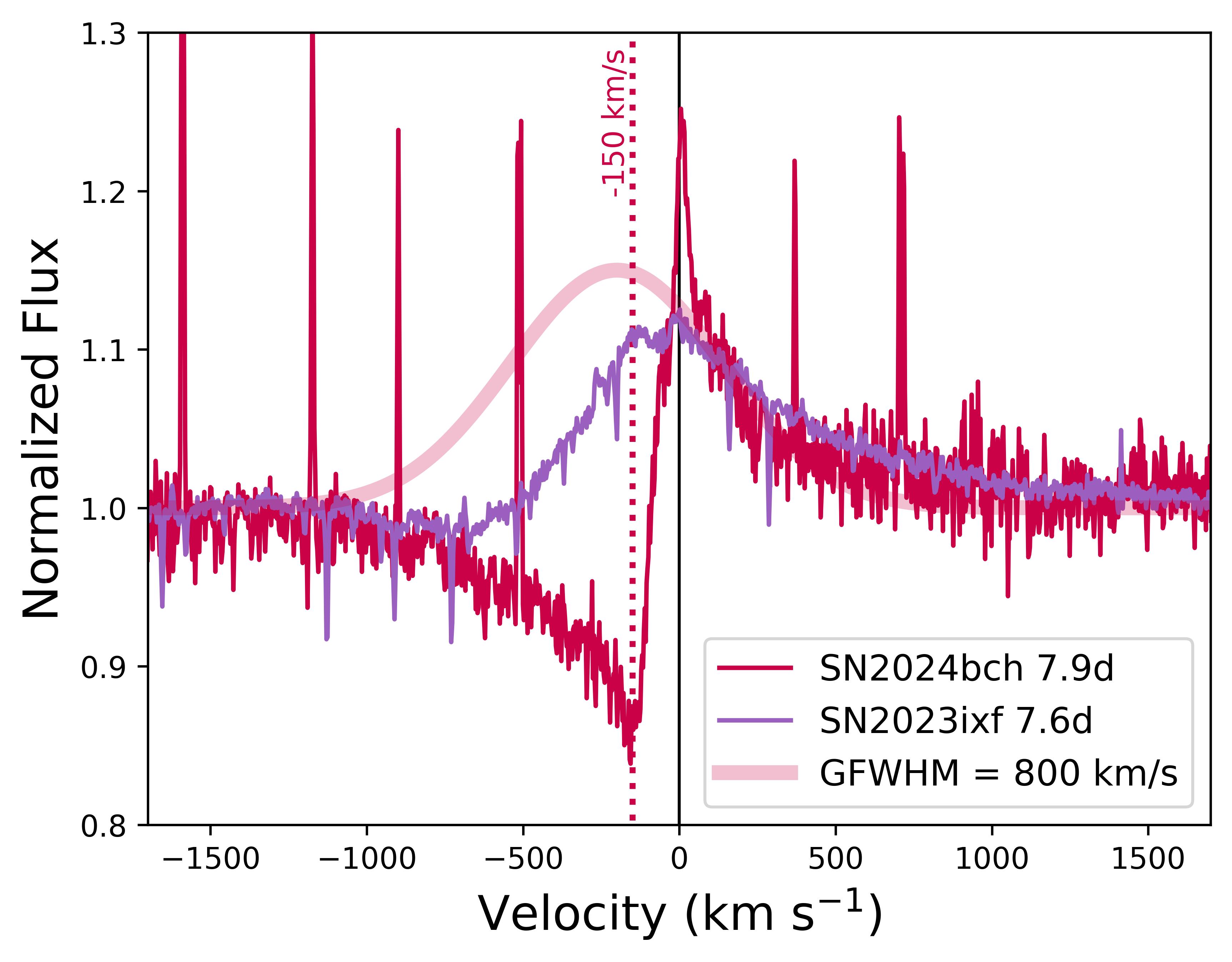}
\includegraphics[width=\linewidth]{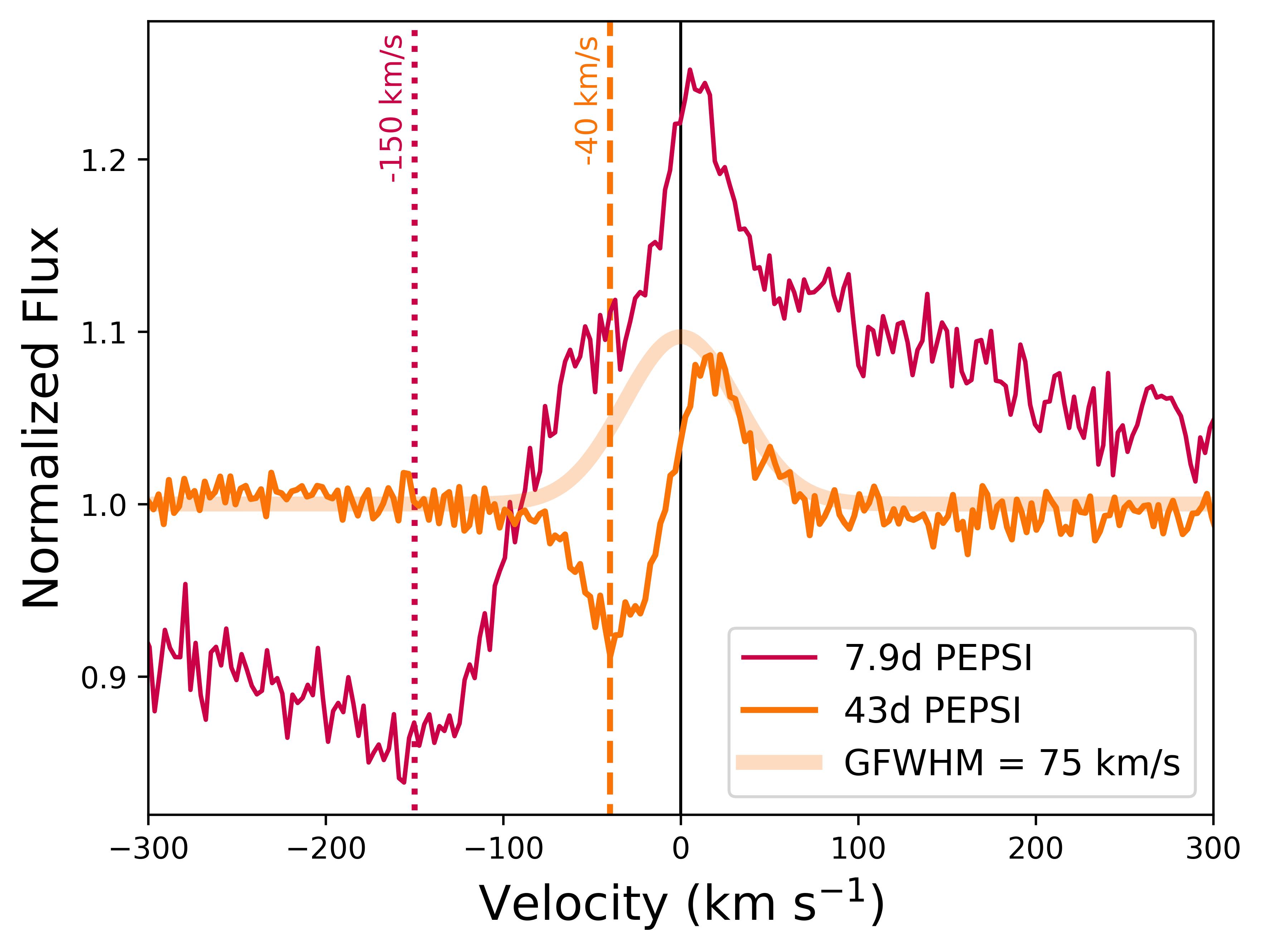}
\caption{Top: GMOS (\J{2.8d}) and PEPSI (\J{7.9d}) spectra of SN~2024bch around H$\alpha$. Middle: Comparison between PEPSI spectra of SN~2024bch on day \J{7.9} and SN~2023ixf on day 7.6 (from \citealt{2023ApJ...956...46S}). Bottom: Day \J{7.9} and 43 PEPSI spectra of SN~2024bch. All spectra have been normalized to the local continuum, and the extremely narrow absorption and emission lines seen in the PEPSI data are telluric in nature.  }
\label{fig:hahires}
\end{figure}

As we show in Figure \ref{fig:hahires}, the medium (GMOS) and high (PEPSI) resolution spectra around H$\alpha$ taken on \J{2.8, 7.9}, and 43 days after explosion give us critical information about the evolution of the ejecta and the physical characteristics of the surrounding CSM. On day \J{2.8} H$\alpha$ can be fit with a Lorentzian profile with a Lorentzian full-width half-max (LFWHM) = 1700~km~s$^{-1}$ centered at $-95$ km s$^{-1}$ with electron scattering wings extending out to about 3000 km s$^{-1}$.  Additionally, there is a small bump just to the blue side of the narrow H$\alpha$ that is emission from \ion{He}{2} $\lambda$6560 \citep[see][]{2023ApJ...956...46S}. The narrow H$\alpha$ is centered at 35 km s$^{-1}$ and has a FWHM = 100 km s$^{-1}$, with a possible P-Cygni absorption at $-60$ km s$^{-1}$ (although with the close \ion{He}{2} it is difficult to tell). This narrow peak is presumably from radiatively heated CSM surrounding the SN that has yet to be hit by the SN shock, and the absorption minimum likely traces the progenitor wind speed. The resolution of the GMOS spectra is roughly 60 km s$^{-1}$, so we are probably not fully resolving the line and overestimating the wind velocity.  Of note is that typical RSG outflow speeds are only about 20 km s$^{-1}$, so the speeds measured here are considerably faster.

 In the day \J{7.9} PEPSI spectrum we see two different components, neither exhibiting the Lorentzian shape as seen in early epochs.  There is now an intermediate width feature that can be fit with a Gaussian having a FWHM = 800 km s$^{-1}$ centered at $-200$ km s$^{-1}$  and showing a very prominent P-Cygni absorption component (as seen in the middle panel of Figure \ref{fig:hahires}), and an additional narrow H$\alpha$ emission feature with a FWHM = 35 km s$^{-1}$ centered at approximately 10 km s$^{-1}$.  This narrow feature may have a very weak P-Cygni absorption with a minimum around $-35$ km s$^{-1}$, but it is at the noise level of the spectrum and hard to disentangle. Assuming it is a real feature, this would put the progenitor wind speed at 35 km s$^{-1}$, much closer to RSG wind speeds, and similar to that found for SN~2024ggi \citep{2024ApJ...972L..15S} \J{which was also observed with high resolution spectroscopy}. Interestingly, this is much slower than the CSM speed measured a few days after explosion in \J{the PEPSI spectrum of} SN~2023ixf \citep{2023ApJ...956...46S}.  If the \ion{He}{2} $\lambda$6560 emission line is still present, it has been eroded by the strong H$\alpha$ absorption. The broader P-Cygni absorption likely arises from gas that has been swept up by the shock, hence the speeds of up to 800--1000 km s$^{-1}$, and cooled. This is also referred to as the cool dense shell (CDS).   This absorption feature has a sharp minimum at $-200$ km s$^{-1}$ and wings extending out to roughly $-1000$ km s$^{-1}$. 
 
 In the bottom panel of Figure \ref{fig:hahires} we zoom in to compare the narrow components in the day \J{7.9} and 43 PEPSI spectra of SN~2024bch. By day 43  the faster, broad absorption line has disappeared, but the narrow emission line is now very pronounced with an obvious and deep P-Cygni absorption. The day 43 narrow line can be fit with a Gaussian with FWHM = 75 km s$^{-1}$ centered at $0$ km s$^{-1}$. The narrow emission component is coming from photoionized CSM, and the P-Cygni absorption, which has a minimum around $-40$ km s$^{-1}$ and extends out to roughly 100 km s$^{-1}$, comes from the unshocked CSM along our line of sight. Similar features have been seen in interacting IIn supernovae observed in high resolution.  For instance \citet{2020MNRAS.499.3544S} show that the IIn SN~2017hcc has narrow absorption persisting at least up to a year after explosion created from the non-shocked CSM.  The later epochs of SN~2017hcc even reveal broader P-Cygni absorption as a result of the illumination of a wider range of absorption speeds along the line of sight, similar to what we see between day 8 and 43 in SN~2024bch. If we assume an average ejecta velocity of 7000 km s$^{-1}$ as above, by 43 days the unshocked CSM would have to be at least 2.6 $\times$ 10$^{15}$ cm away, and assuming a progenitor wind velocity of 35 km s$^{-1}$ from the width of the narrow line, it would have to have been ejected about 25 years before explosion. This is an order of magnitude larger than the value estimated from the disappearance of the high ionization lines above. This discrepancy could be due to two distinct areas of CSM, one close in from the most recent mass loss of the progenitor creating the early narrow lines, and another extended region further out with possible asymmetries in the form of a disc or torus. The fact that we do see the narrow absorption features does suggest that at least some portion of the density enhancement if there is one, is along our line of sight.

In the middle panel of Figure \ref{fig:hahires} we also compare PEPSI data of SN~2024bch on day \J{7.9} with the PEPSI data of SN~2023ixf on day 7.6 from \citet{2023ApJ...956...46S}.  While both SNe have similar red wing H$\alpha$ emission, and the Gaussian FWHM for SN~2024bch is similar to the 900 km s$^{-1}$ found for SN~2023ixf, the P-Cygni absorption is much broader and deeper in SN~2024bch, and spans a much larger range of velocities. This could indicate a different CSM geometry for the two objects. \citet{2023ApJ...956...46S} concluded that the CSM of SN~2023ixf was indeed asymmetric but was likely engulfed by the SN photosphere. While the possible asymmetric nature of the CSM for SN~2024bch will be discussed in more detail in section \ref{sec:asym}, the fact that we still see narrow P-Cygni H$\alpha$ lines on day 43 does suggest that the CSM, at least along our line of sight, has not been completely swept up by the SN ejecta. If the CSM is asymmetric, it also requires that our line-of-sight has a vantage point looking through the dense CSM.

\subsection{H$\alpha$ Evolution}
\label{sec:halpha}
\begin{figure}
\includegraphics[width=\linewidth]{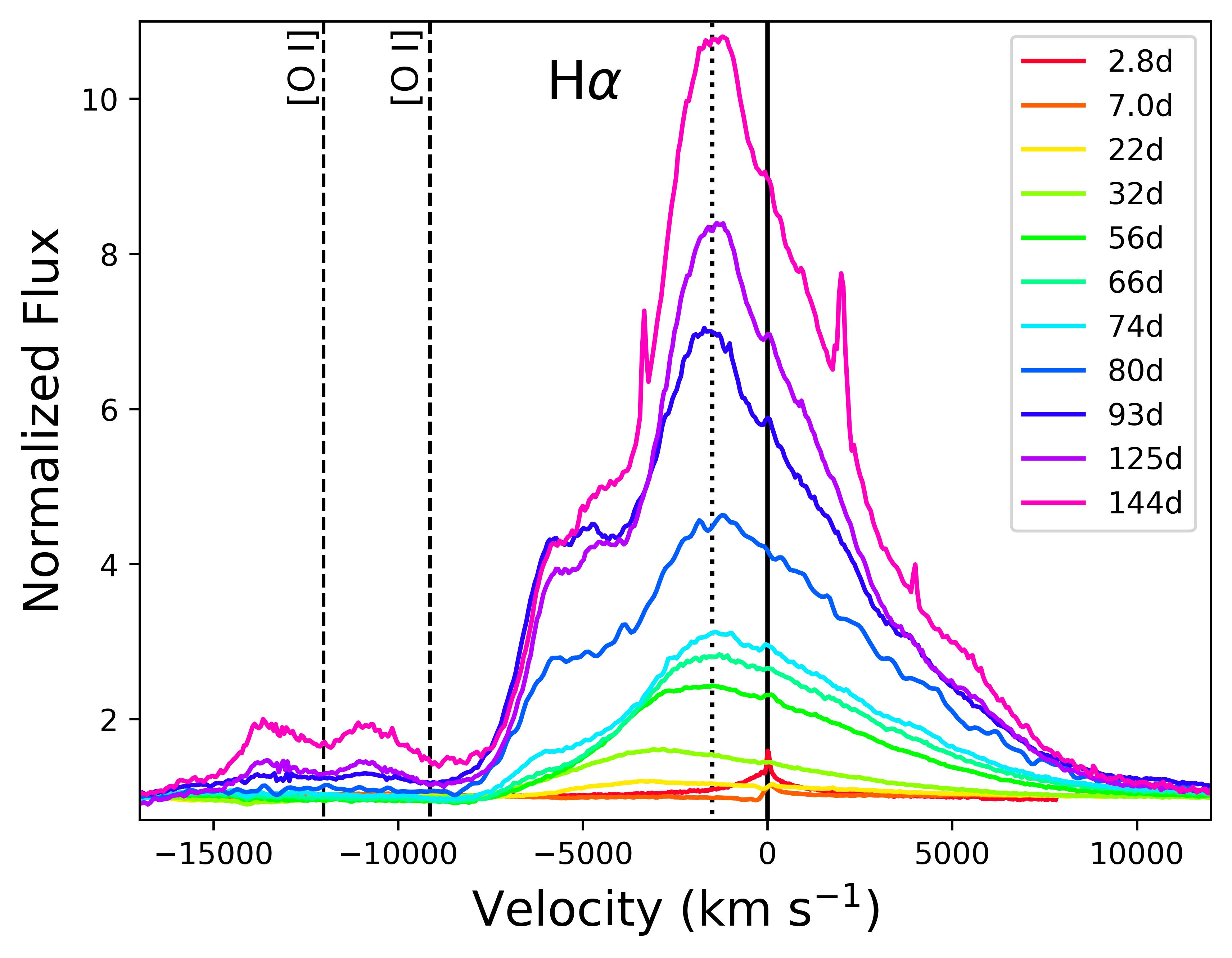}
\caption{H$\alpha$ evolution in velocity space for SN~2024bch. The dotted line marks $-1500$ km s$^{-1}$, where the blueshifted peak settles after the first month.  A multi-component feature begins to emerge after day 50 around  $-1500$ km s$^{-1}$, and by day 144, it has eroded and migrated redward. The location of the forbidden [O~I] lines at rest velocity have also been marked with dashed lines. They too have blueshifted peaks at $-1500$ km s$^{-1}$. }
\label{fig:haevolution}
\end{figure}

In Figure \ref{fig:haevolution} we show the evolution of H$\alpha$ in our lower resolution mostly Gemini/GMOS spectra from $\sim$ day 3 to day 144 normalized to the local continuum. While the early spectra show the narrow H$\alpha$ discussed above, by day 22 a broad ($\sim$7000  km s$^{-1}$), blueshifted emission line starts to emerge. Blueshifted broad H$\alpha$ lines are expected in the early spectra of Type II SNe as the redshifted receding side of the line is obscured by the opaque hydrogen envelope \citep{2005A&A...439..671D,2014MNRAS.441..671A}. The blueshifted H$\alpha$ peak migrates from $-3200$ km s$^{-1}$ to $-1500$ km s$^{-1}$ by day 56 where it stays until our last spectrum before Sun-occultation. Generally we expect the H$\alpha$ line peak to move to zero velocity after the photospheric phase ends \citep{2014MNRAS.441..671A}, but this does not seem to be the case for SN~2024bch, at least by day 147.  This could possibly be due to asymmetries in the supernova ejecta, an assumption further supported by the blueshifted peaks of the emerging [\ion{O}{1}] doublet also at $-1500$ km s$^{-1}$. The Type II SNe 2007it, 2017gmr, and more recently 2023ixf, all showed H$\alpha$ and [\ion{O}{1}] emission blueshifted by the same amount   \citep{2011ApJ...731...47A,Andrews2019,2024A&A...687L..20F}. For the cases of SN~2017gmr and SN~2023ixf, spectropolarimetric observations further suggest intrinsic asymmetries in the SN ejecta \citep{2019MNRAS.489L..69N,2024arXiv240520989S,2023ApJ...955L..37V,2024arXiv241008199S}. This could very well be the case for SN~2024bch as we will explore more below.

Starting around 60--70 days, as the SN transitions from the plateau to the radioactive decay tail, an additional H$\alpha$ emission feature begins to emerge around $-6000$ km s$^{-1}$. Between day 80 and 125 this feature becomes very prominent and a bit flat topped, spanning between $-4000$ and $-6000$ km s$^{-1}$, but by day 144 it has eroded and the center has moved redward to  $-4500$ km s$^{-1}$.  In this last epoch, it is also possible that a similar feature is emerging on the red side around 4500 km s$^{-1}$.  Unfortunately, the SN becomes Sun constrained after this epoch, but future observations will be able to track this evolution. 
In the Type II SN PTF11iqb which also shows multi-peaked H$\alpha$ and an emergence and eventually domination of a red side component, \citet{PTF11iqb} suggested that the asymmetric CSM was enveloped by the ejecta and was uncovered again as the photosphere receded. A similar situation could be occurring in SN~2024bch.
Multi-component H$\alpha$ like SN~2024bch exhibits has been observed in numerous other Type II SNe as well, and the postulated reason for these features is generally interaction with surrounding asymmetric CSM or an intrinsic asymmetry in the SN ejecta/explosion itself \citep[for example]{2005ApJ...622..991F,2017MNRAS.471.4047A,Andrews2019,2021MNRAS.505..116U,2023MNRAS.523.5315P}.   We will explore this more in depth in section \ref{sec:asym}.

\begin{figure*}
\includegraphics[width=\linewidth]{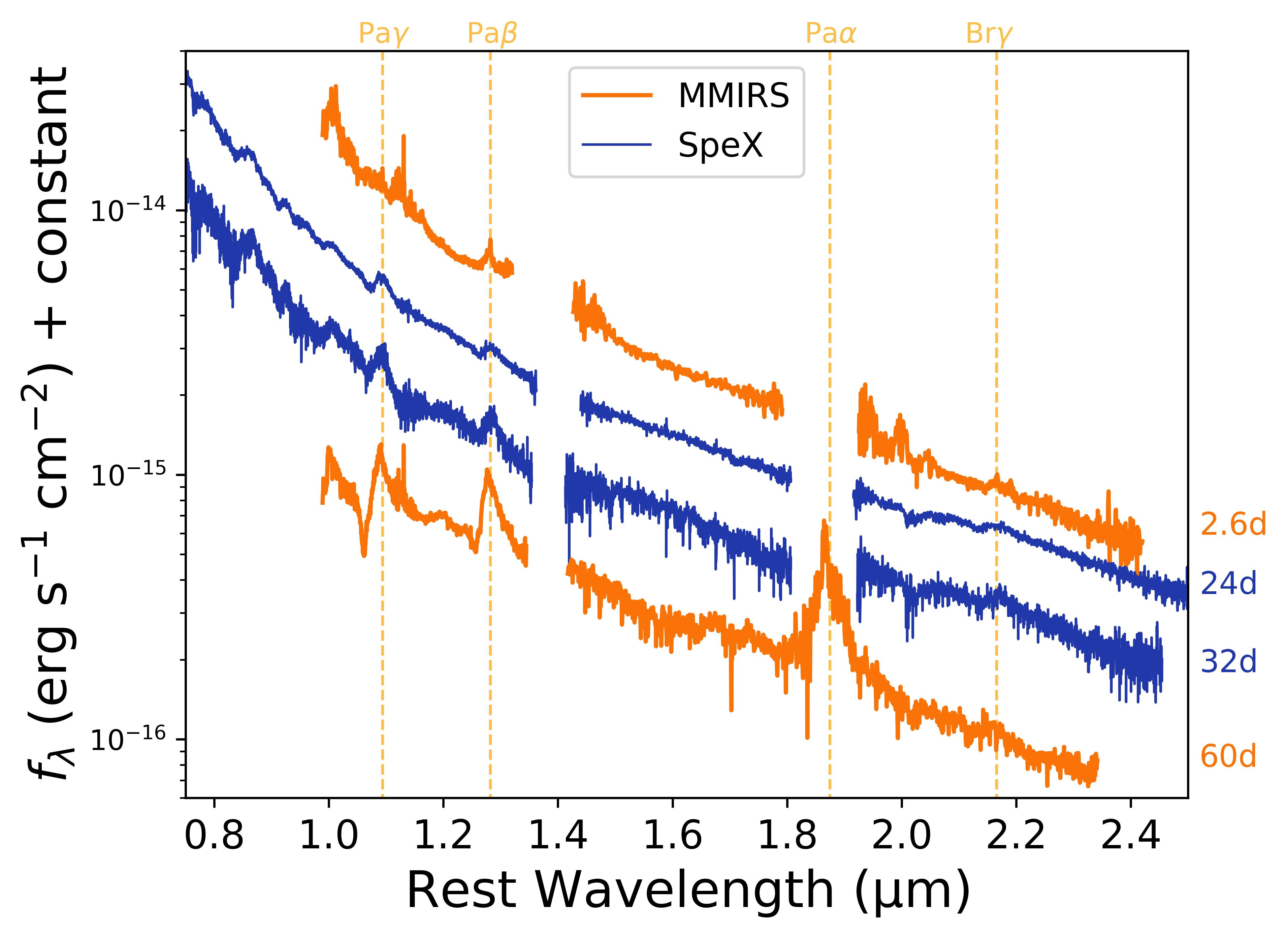}
\caption{NIR spectra of SN~2024bch as described in Table \ref{tab:nirspec}. Notable lines are identified, and the dates are with respect to our assumed explosion epoch of MJD \J{60337.65}.}
\label{fig:nirspec}
\end{figure*}

\subsection{NIR Spectra}
\label{sec:NIR}
\begin{figure}
\includegraphics[width=\linewidth]{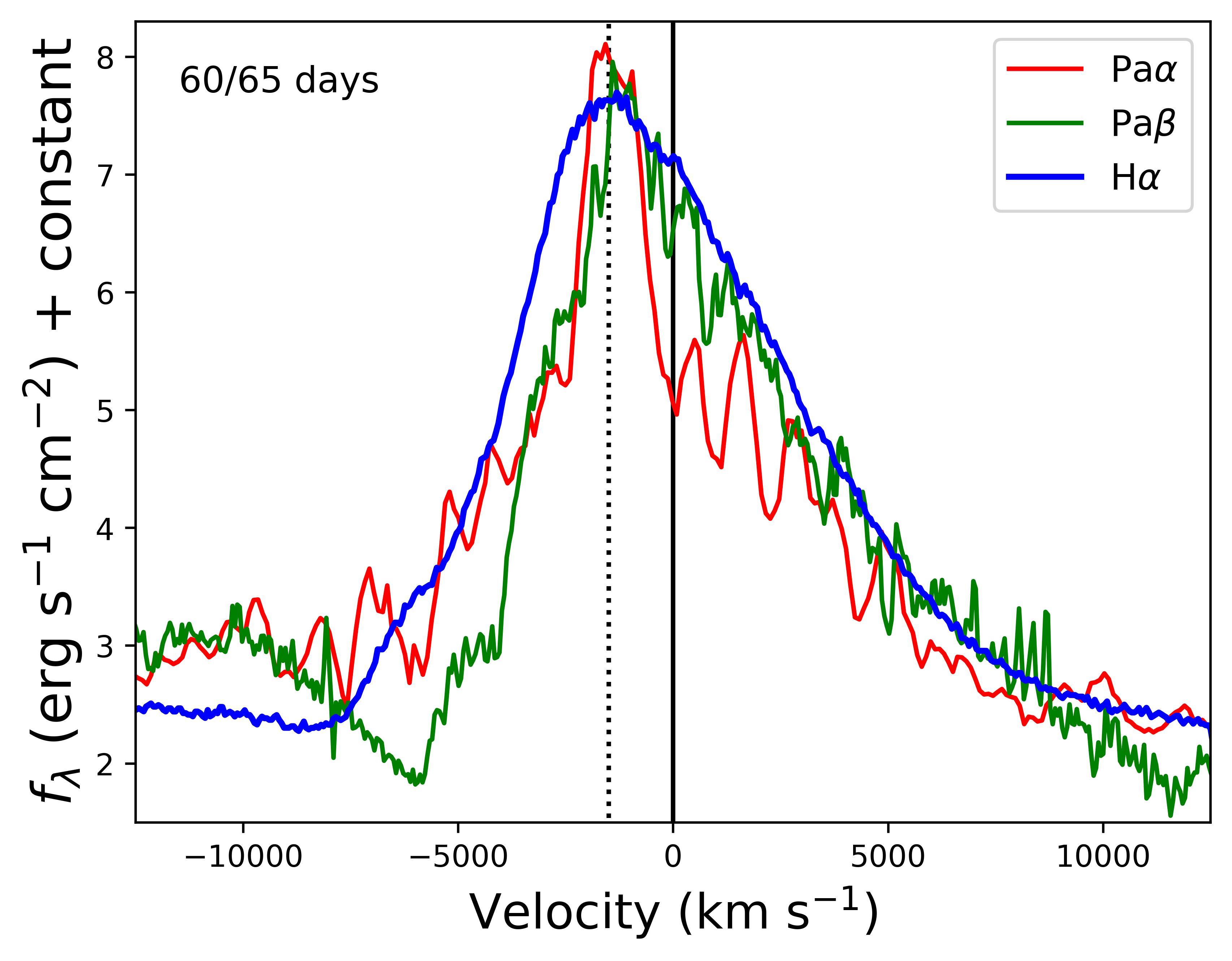}
\caption{Hydrogen evolution in velocity space for SN~2024bch on day 60 (NIR) and 65 (optical). The dotted line marks $-1500$ km s$^{-1}$, where all lines peak.   A multi-component feature is noticeable at $-5000$ km s$^{-1}$ in Pa$\beta$, as is a feature at $-6000$ km s$^{-1}$ in H$\alpha$.}
\label{fig:day60spec}
\end{figure}

Figure \ref{fig:nirspec} shows the NIR spectral evolution from \J{2.6} to 60 days after explosion. The first spectrum is mainly featureless, except for narrow Pa$\beta$ and weak Br$\gamma$ emission. Flash emission in the NIR is not commonly observed, whether due to lack of early spectra or lack of actual narrow lines.  For example, in SN~2017gmr the NIR spectrum taken on day 2.8 was completely featureless \citep{Andrews2019}.  In the following weeks, Pa$\alpha$ broadens, and Pa$\gamma$ emerges.  By our last spectrum on day 60 strong Pa$\alpha$ is seen, and all Paschen lines we observe have blueshifted peaks at $-1500$ km s$^{-1}$.  In Figure \ref{fig:day60spec} we show Pa$\alpha$, Pa$\beta$, and H$\alpha$ in velocity space around day 60.  The blueshifted peak is clear, and an additional component around $-5000$ km s$^{-1}$ is seen emerging in Pa$\beta$, a week or so before it becomes obvious in H$\alpha$ as discussed above. 

Dust formation could be invoked for a blueshift in an emission line, but the NIR spectra of SN~2024bch do not show the prominent CO overtone band between 2.3-2.5 $\micron$ that is generally seen as a precursor to dust formation \citep{2015A&A...575A..95S}. CO has been seen in quite a few CCSNe including SN~2004et \citep{2005ApJ...628L.123K}, SN~2013ej \citep{2016MNRAS.461.2003Y}, and SN~2017eaw \citep{2018ApJ...864L..20R,2019ApJ...873..127T}. Additionally, Figure  \ref{fig:day60spec} shows that there is no wavelength dependent attenuation, further supporting that dust formation could not be the reason for this line asymmetry. It is possible the CO feature could show up at later epochs in SN~2024bch when the ejecta has cooled sufficiently for grains to form, so continued observations in the IR are prudent.

\begin{figure*}
\includegraphics[width=\columnwidth]{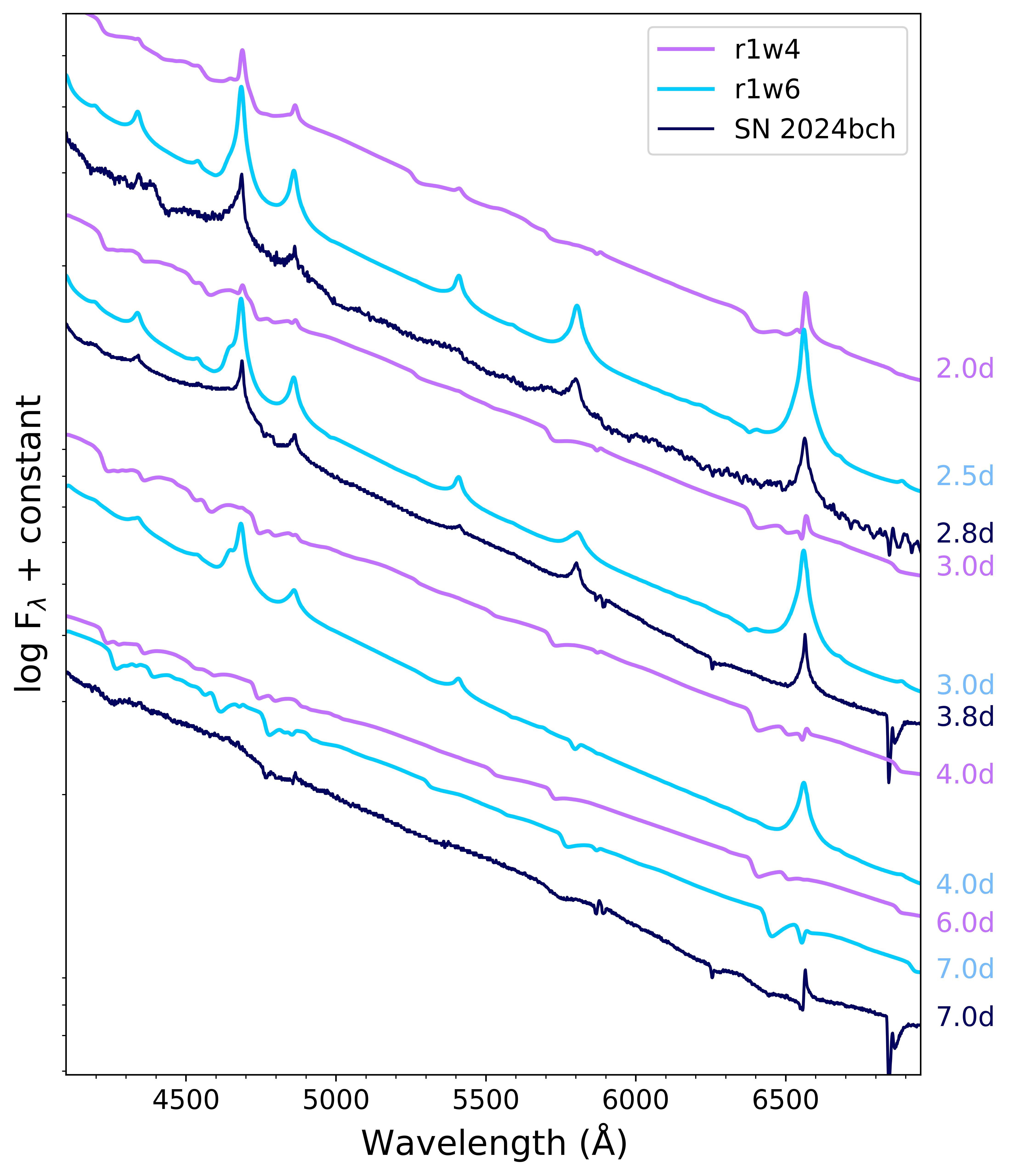}
\includegraphics[width=\columnwidth]{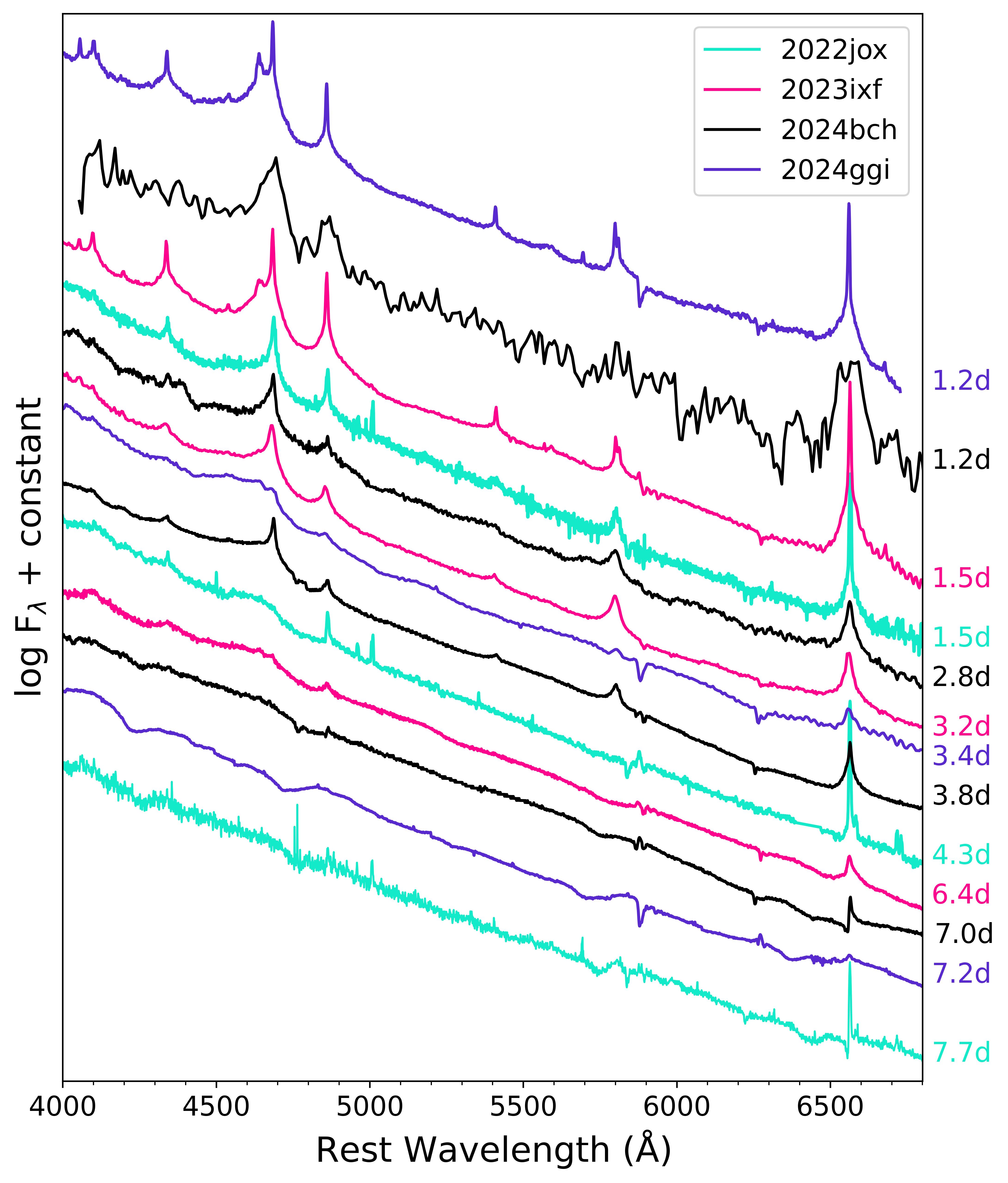}
\caption{Left: Comparison of the early optical spectra of SN~2024bch to the \texttt{r1w4} and \texttt{r1w6} models of \citet{Dessart17}. These models have a mass loss rate of 10$^{-3}$ M$_{\sun}$ yr$^{-1}$ and 10$^{-2}$ M$_{\sun}$ yr$^{-1}$ respectively. Right: Comparison of early spectra of SN~2024bch to similar features seen in SN~2022jox \citep{2024ApJ...965...85A}, SN~2023ixf \citep{2023ApJ...956L...5B}, and SN~2024ggi \citep{2024ApJ...972L..15S}.}
\label{fig:DessartComp}
\end{figure*}

\section{Analysis}
\label{sec:analysis}
\subsection{Model Comparison at Early Times}

In Figure \ref{fig:DessartComp} we show comparisons of early SN~2024bch spectra with non-LTE radiative transfer models from \citet{Dessart17}. These models assume a 15 M$_\sun$ RSG with $R_{\star} = 501\ R_{\sun}$ with various mass loss rates and span the time period from shock breakout to roughly 15 days post explosion.  The wind velocity, $v_{w}$ = 50 km s$^{-1}$, extends to 5 $\times$ 10$^{14}$ cm at which point the mass loss rate drops to 10$^{-6}\ M_\sun\ \mathrm{yr}^{-1}$. Here we show the models of \texttt{r1w4} and \texttt{r1w6}, which have $\dot{M} = 10^{-3}\ M_\sun\ \mathrm{yr}^{-1}$ and $\dot{M} = 10^{-2}\ M_\sun\ \mathrm{yr}^{-1}$, respectively. See \citet{Dessart17} their Table 1 for the parameters of each publicly available model. One thing to note is that each of these models are created for different age intervals, and the sampling of the SN~2024bch optical spectra is not particularly high during this time, so a direct comparison between the observed spectra and the models is not always possible.

As seen in Figure \ref{fig:DessartComp}, for SN~2024bch the persistence of the narrow Balmer, \ion{He}{2}, and \ion{C}{4} lines until day 3.8 and likely beyond, and the overall shape of the continuum slope is most similar to the high mass-loss rate model \texttt{r1w6} which uses an $\dot{M} = 10^{-2}\ M_\sun\ \mathrm{yr}^{-1}$.  At no time do we see \ion{N}{3}/\ion{C}{3} emission just blueward of \ion{He}{2} $\lambda$4685 that is seen in many other SNe with flash spectra (right panel of Figure \ref{fig:DessartComp} for example). It is very likely that the line was present but has faded by this epoch (see \citet{2024arXiv240915431T}). The very low resolution (R$\sim$100) spectrum retrieved from TNS from day \J{1.2} \citep{2024TNSCR.284....1B} does not have the resolution sufficient enough to differentiate individual lines in that region. Overall, the model comparison suggests that the mass-loss rate for SN~2024bch's progenitor was likely between $\dot{M} = 10^{-3} - 10^{-2}\ M_\sun\ \mathrm{yr}^{-1}$. 
 
We have also compared our \J{2.8} day FLOYDS spectrum with models from \citet{2019A&A...621A.109B} that simulate SN ejecta interaction with CSM roughly a day after explosion. They model three different progenitor types with different surface abundances: a solar abundance representing a low mass RSG, CNO-processed abundances from higher mass red-, blue-, and yellow- supergiants, and He-rich abundances corresponding to stars like LBVs or Wolf-Rayet stars.  In the SN~2024bch spectrum we do not see any \ion{O}{6}, [\ion{O}{3}], \ion{N}{3}, or \ion{C}{3}, but there is fairly strong \ion{H}{1}, \ion{C}{4}, and \ion{He}{2} $\lambda$4685 emission. There are also weak but observable \ion{He}{2} $\lambda5412,6559.8$ and \ion{N}{4} lines. As mentioned before, it is possible that some lines may have been present earlier and have faded by this epoch.

\begin{figure}
\includegraphics[width=\columnwidth]{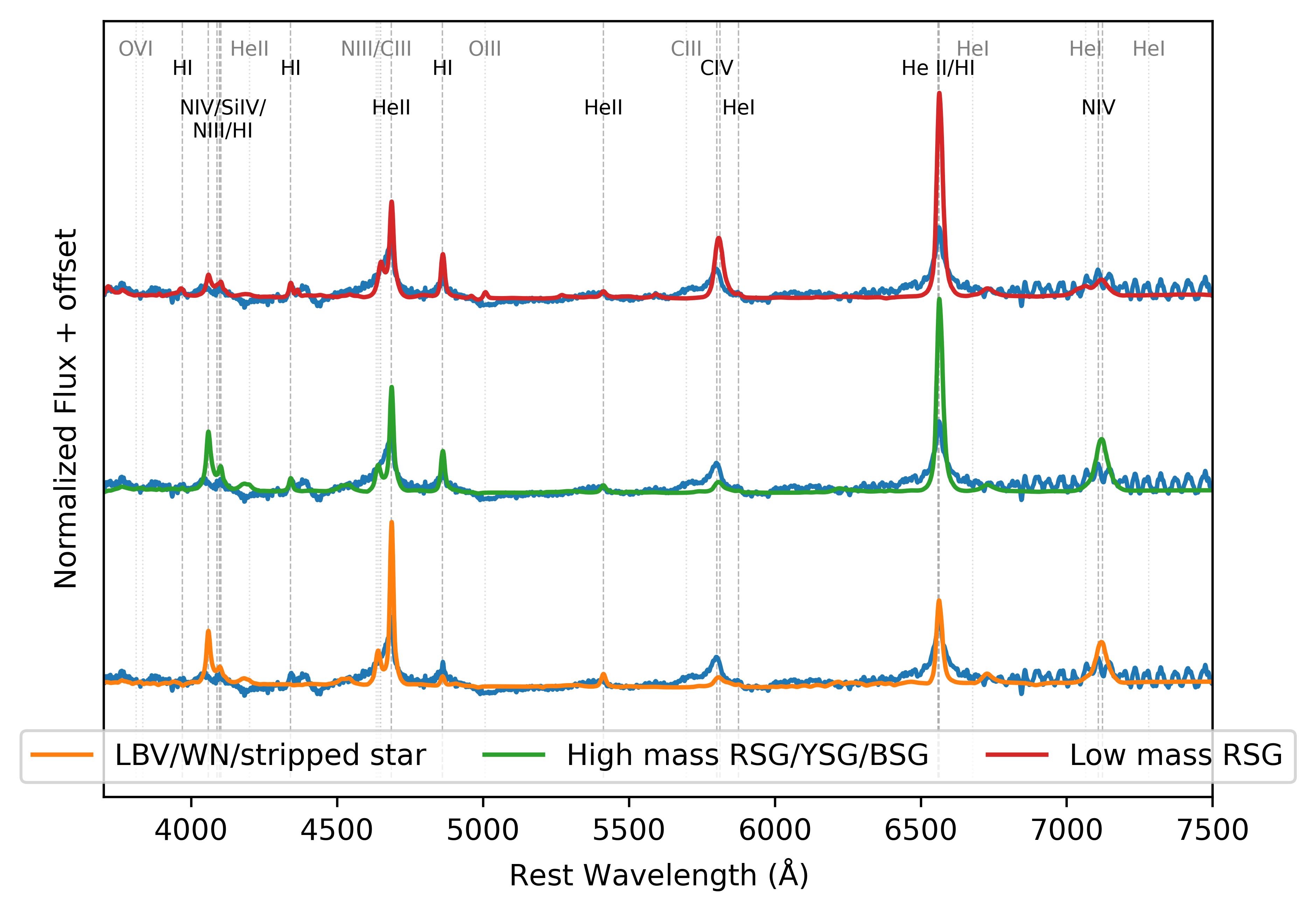}
\includegraphics[width=\columnwidth]{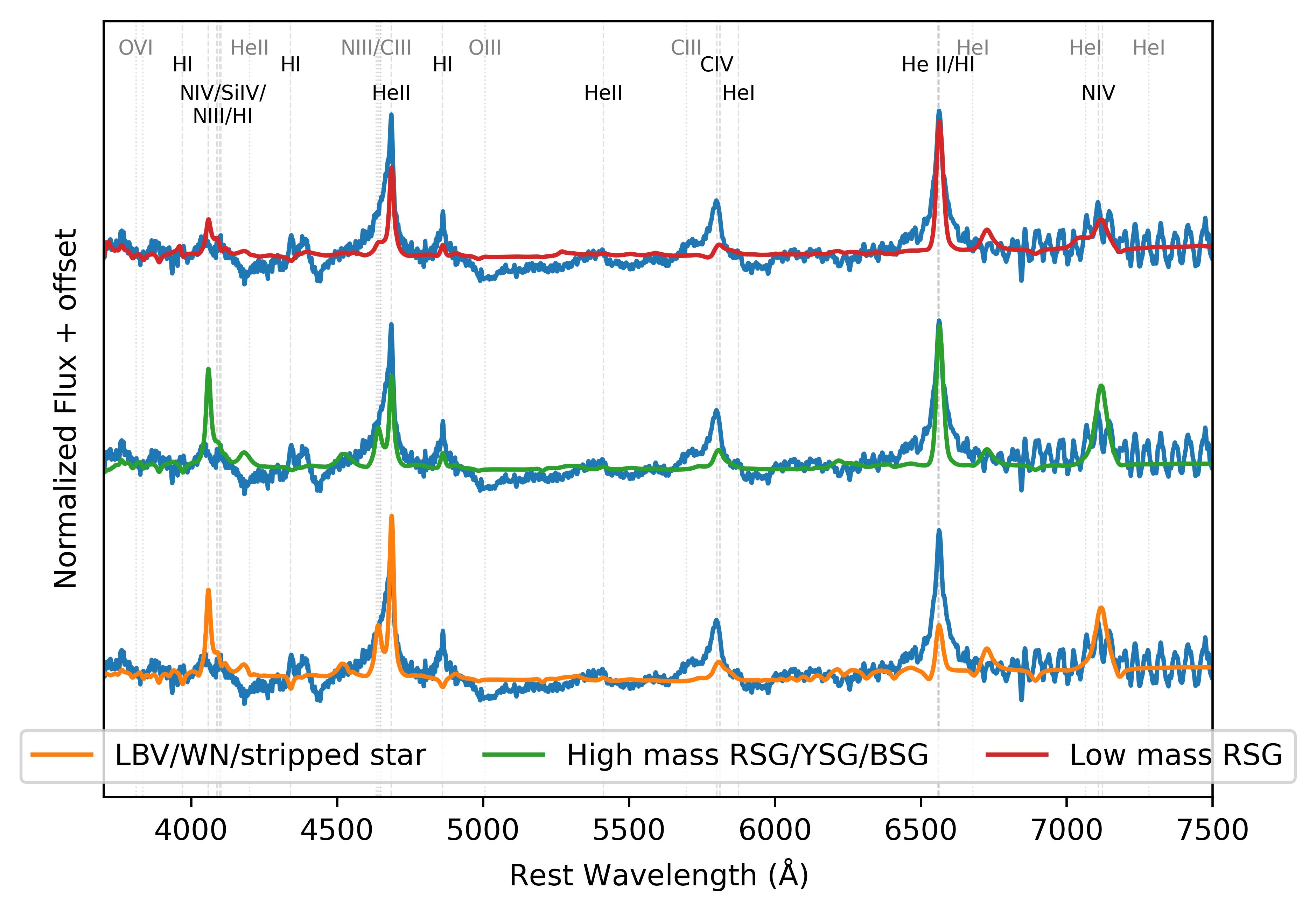}
\caption{The day 2.8 FLOYDS spectrum (blue) of SN~2024bch compared with L = 1.5 $\times$ 10$^{9}$ L$_{\odot}$ models from \citet{2019A&A...621A.109B}. The model on the top is for an $\dot{M}$ = 1.0 $\times$ 10$^{-3}\ M_\sun\ \mathrm{yr}^{-1}$, and on the bottom for $\dot{M}$ = 5.0 $\times$ 10$^{-4}\ M_\sun\ \mathrm{yr}^{-1}$. Red lines are solar abundance models, green lines are CNO-processed abundances, and orange lines are from He-rich abundance models. Dark gray lines identify emission lines seen in our spectra, while the light gray lines are lines present in the model but not in the observed spectrum. The models have been convolved with a Gaussian kernel to more closely match the native resolution of the spectrum. }
\label{fig:BoianComp}
\end{figure}

As we show in Figure \ref{fig:BoianComp} we can manage a decent match between SN~2024bch and models with L = 1.5 $\times$ 10$^{9}$ L$_{\odot}$ and mass loss rates of $\dot{M}$ = 1.0 $\times$ 10$^{-3}\ M_\sun\ \mathrm{yr}^{-1}$ (left) and $\dot{M}$ = 5.0 $\times$ 10$^{-4}\ M_\sun\ \mathrm{yr}^{-1}$ (right). The models are created at high resolution \citep{2019A&A...621A.109B}, so they have been convolved with a Gaussian kernel so that the resolution matches the lower resolution of the data. The higher mass-loss models over-estimate the H$\alpha$ and \ion{He}{2}, and the RSG and hot supergiant star models in particular both show the \ion{N}{3}/\ion{C}{3} emission not seen in our spectra. In the lower mass-loss rate model on the bottom there is a decent match with the intensity of the line emission in the data, but the \ion{C}{4} is very weak in the model, and the Balmer and \ion{He}{2} lines are not completely reproduced by any one model.

Our earliest spectra were taken at \J{2.8} days after estimated explosion, almost two days after the models, so the comparison needs to be taken with some caution. Using the scaling relation as described in \citet{2019A&A...621A.109B} though, we can calculate an L = 2.0 $\times$ 10$^{9}$\ L$_{\odot}$, and a range of $\dot{M}$ = 3 -- 7 $\times$ 10$^{-3}\ M_\sun\ \mathrm{yr}^{-1}$.  This mass loss rate agrees fairly well with the comparison with the models from \citet{Dessart17} discussed above, and is similar to estimated mass-loss rates from other Type II SNe as we will discuss in Section \ref{sec:comp}. 

\subsection{Comparison with other Flash SNe}
\label{sec:comp}

Over the past few years the sample size of CCSNe with extremely early and high-cadence observations in the hours and days after explosion has grown substantially. In the last year alone, two nearby ($<$ 7 Mpc) SNe, 2023ixf and 2024ggi were discovered and followed within a day of explosion, and both showed flash ionization lines.  This makes them great objects to compare with SN~2024bch.  While we show the $V$-band lightcurve comparison in Figure \ref{fig:colorcompare} and the $B-V$ comparison in Figure \ref{fig:colorflash}, on the right side of Figure \ref{fig:DessartComp} we also show the comparison of the early optical spectra and include another recent Type II SN with good early coverage \J{and flash features,} SN~2022jox.  The data come from \citet[SN~2022jox]{2024ApJ...965...85A}, \citet[SN~2023ixf]{2023ApJ...956L...5B}, and \citet[SN~204ggi]{2024ApJ...972L..15S}. 

Unfortunately we do not have spectra with suitable resolution before \J{2.8} days for SN~2024bch, but there are obvious similarities in \ion{He}{2} $\lambda$4685, \ion{C}{4}, and the Balmer lines between SN~2024bch and SN~2023ixf during this epoch. Similarly, there are no SN~2022jox spectra around day 3, but the day 1.5 spectrum also looks very similar to SN~2024bch. We also lack spectra for SN~2024bch between day \J{3.8 and 5.9} (shown in Figure \ref{fig:earlyspec}) so it is unclear when the \ion{He}{2}  and \ion{C}{4} lines fully disappear, but SN~2023ixf and SN~2024bch seem quite similar around day 7 as well, with the notable exception being the obvious strong narrow P-Cygni absorption seen in H$\alpha$ for SN~2024bch. This absorption feature is also seen in SN~2022jox 7.7 days after explosion. The flash features in SN~2024ggi evolve quickly, and are mostly gone (minus some Balmer line emission) by day 3 \citep{2024ApJ...972L..15S}, and overall do not match temporally with the other three SNe. Similarly, the $V$-band lightcurves of SN~2022jox, SN~2023ixf, and SN~2024bch are all fairly similar, while SN~2024ggi has an extended plateau (Figure \ref{fig:colorcompare}), and as shown in Figure \ref{fig:colorflash} and mentioned in \citet{2024ApJ...972L..15S}, the $B-V$ color of SN~2024ggi also reaches a maximum blue color much earlier than the other three.  

Mass-loss estimates for SN~2022jox and SN~2023ixf are all between $\dot{M} = 10^{-3} - 10^{-2} \ M_\sun\ \mathrm{yr}^{-1}$  based on the analysis of the early spectra \citep{2024ApJ...965...85A,2023ApJ...954L..42J,2023ApJ...955L...8H,2023arXiv230610119B,2023SciBu..68.2548Z} which is consistent with the model fits to SN~2024bch described above. The measured $L = 2.0 \times 10^{9}\ L_{\odot}$ using \citet{2019A&A...621A.109B} is also similar to the measured luminosites using the same method for SN~2024ggi \citep[$2.9 \times 10^{9}\ L_{\odot}$,][]{2024ApJ...972L..15S} and SN~2023ixf \citep[$2.6 \times 10^{9}\ L_{\odot}$,][]{2023ApJ...956L...5B}. Interestingly even though SN~2024ggi has mass-loss rate estimates similar to the other SNe, signs of flash interaction fade more quickly (see Figure \ref{fig:colorflash}).  This range of flash duration but with similar mass-loss rates could point to different mass-loss histories and various geometries and densities of the surrounding CSM.  

Finally, if we compare the high resolution spectra of SN~2024bch to that of SN~2023ixf presented in \citet{2023ApJ...956...46S} there are some interesting differences, even though other physical characteristics may look similar.  In SN~2023ixf there was a lack of any narrow P-Cygni absorption seen in the echelle spectra that was clearly seen in SN~2024bch. This could be explained fairly easily by asymmetric CSM where our line of sight is closer to edge on for SN~2024bch than SN~2023ixf.  What is more puzzling in the fairly broad emission from the pre-shock CSM traced by the H$\alpha$ emission in SN~2023ixf that is not see in SN~2024bch (see center panel in Figure \ref{fig:hahires}). While we could invoke radiative acceleration for the broadening seen in SN~2023ixf, the luminosities of the two objects are quite similar (see the lightcurve comparison in Figure \ref{fig:colorcompare} and previous paragraph) and therefore we would expect a similar feature in SN~2024bch.  This could suggest that the radiative acceleration is not the main reason for the broadened components, but that the geometry may play the primary role as discussed in \citet{2023ApJ...956...46S}.

\subsection{Asymmetries in ejecta or CSM}
\label{sec:asym}

Multiple features of the optical spectra of SN~2024bch lend themselves to the interpretation of asymmetries in either the CSM and/or the SN ejecta.  In particular the blueshifted peaks of the Balmer lines along with those of [\ion{O}{1}] in the nebular phase suggest that the explosion itself was asymmetric. ``Horned" [\ion{O}{1}] emission line profiles, where symmetric peaks are distributed around zero velocity, have been thoroughly studied in stripped envelope supernovae (SESN) \citep{2008ApJ...687L...9M,2009MNRAS.397..677T,2022ApJ...928..151F,2024NatAs...8..111F} and is attributed to aspherical ejecta geometries. These features seem to be less common in hydrogen rich SNe, but there are a few cases in the literature. Similar features were seen in SN~2023ixf \citep{ 2024A&A...687L..20F} and SN~2017gmr \citep{Andrews2019}. These objects also had corresponding spectropolarimetry observations that confirm the asymmetric nature of the explosion \citep{2019MNRAS.489L..69N,2023ApJ...955L..37V,2024arXiv240520989S,2024arXiv241008199S}.  In particular, \citet{2021MNRAS.505..116U} and \citet{2024arXiv240903540F} suggest that bipolar or toroidal oxygen ejecta can adequately reproduce the line profile of [\ion{O}{1}] in SN~2023ixf.

As discussed in Section \ref{sec:NIR}, blueshifted H$\alpha$ profiles can sometimes be explained by dust formation in the ejecta, which will preferentially attenuate the red receding side more so than the blue.  This can be accompanied by a simultaneous increase in the IR brightness and a decrease in the optical lightcurves.  While we do see the bolometric lightcurve fade a bit faster than $^{56}$Co decay, this could also be due to other factors such as incomplete gamma ray trapping or CSM interaction as discussed above.  Dust extinction is wavelength dependent, so if dust formation is occurring in the ejecta we should expect the red side of bluer emission lines to be more attenuated than the red side of the redder emission lines.  If we look at Figure \ref{fig:day60spec}, the shapes of H$\alpha$ and Pa$\alpha$ are similar, particularly on the red side, suggesting that new dust formation is likely not the cause for this persistent blueshift.

The multi-component H$\alpha$ (and other hydrogen lines) seen in SN~2024bch may require additional asymmetries in the CSM. Asymmetric CSM interaction is not uncommon \citep{2000ApJ...536..239L,PTF11iqb,2010ApJ...715..541A,2017MNRAS.471.4047A,smith17review,2018MNRAS.477...74A,Andrews2019,2020MNRAS.499.3544S,2023ApJ...956...46S}, and the high binary fraction of massive stars \citep{2012Sci...337..444S} makes it likely that some objects may experience pre-SN binary interaction triggered by envelope inflation \citep{2014ApJ...785...82S,2013arXiv1302.5037S}. The appearance of the additional H$\alpha$ feature around $-5000$ km s$^{-1}$ that starts to emerge around day 60 and persists until our last epoch is likely being formed in the post-shock region between the forward and reverse shocks as the ejecta hits the CSM. That we only see a blue feature could easily be explained by a CSM geometry where from the observer's viewing angle any back-side interaction is obscured by the SN ejecta. This hypothesis is further supported by the fact we still see narrow absorption from unshocked CSM in our nebular spectra.  It is possible that an additional multi-component feature is emerging on the red side at a similar velocity in our last spectrum, but future observations are needed to confirm this.  Of course one sided CSM, caused by various mechanisms could also be invoked, as it has been for previous objects \citep{PTF11iqb,2017A&A...599A.129T,2018MNRAS.475.1104B}.

\section{Summary and Conclusions}
\label{sec:summary}
We have presented comprehensive photometric and spectroscopic data of the Type II SN 2024bch from the UV to IR over the first 150 days. The \J{2.8} day optical spectra shows lines of \ion{H}{1} \ion{He}{2}, \ion{C}{4}, and \ion{N}{4} but lack \ion{C}{3}, \ion{N}{3}, and \ion{O}{4}. By day 6 all but the narrow Balmer lines persist. High-resolution spectra reveal narrow P-Cygni in H$\alpha$ that constrains the progenitor wind to 35--40 km s$^{-1}$, speeds typically associated with RSGs. Comparison between models and the early spectra indicate a mass loss rate of between $\dot{M} = 10^{-3} - 10^{-2} \ M_\sun\ \mathrm{yr}^{-1}$. This requires a period of enhanced mass in the years to weeks before explosion as these values are four orders of magnitude higher than normal quiescent winds from RSGs.

After day 60 when the lightcurve begins to fall from the plateau we also see the emergence of multi-peaked H$\alpha$, likely from the SN shock still interacting with an asymmetric CSM. The lack of a corresponding red component, and the persistence of narrow H$\alpha$ absorption points to CSM in a disc or torus that we are viewing nearly edge-on. This CSM is an order of magnitude further out than the CSM giving rise to early interaction signatures, which hints at a complex CSM geometry and mass-loss history for the progenitor of SN~2024bch. Additionally, the \ion{H}{1} and [\ion{O}{1}] lines all show blueshifted peaks that persist well into the nebular phase, likely from asymmetries in the SN explosion itself.

\J{The interpretation presented here is similar to that of \citet{2024arXiv240915431T} with just one notable difference.  We agree that CSM interaction is evident in the optical spectra after day 60, but also see evidence for interaction at earlier times. This is at odds with their conclusion that CSM interaction does not play a vital role in the early evolution of SN 2024bch, and that Bowen fluorescence may be the source of early narrow lines. While it is true that shock interaction does not need to be invoked to explain the luminosity of SN 2024bch, our spectroscopy, particularly the high-resolution echelle spectra, do suggest early and consistent, albeit possibly low-level, interaction.}

As is becoming more clear through comprehensive studies of CCSNe from hours after explosion until well into the nebular phase, CSM interaction at various levels may be more ubiquitous than once believed. More ``normal" Type II supernovae are showing signs of significant CSM interaction, whether it be via early flash signatures or from late time multi-component H$\alpha$ lines. Continued early and long-term monitoring of these objects is needed to understand the full picture of the CSM and pre-SN mass-loss history of massive stars.

\section*{Acknowledgements}
We thank the anonymous referee for their constructive comments.
J.E.A.\ is supported by the international Gemini Observatory, a program of NSF's NOIRLab, which is managed by the Association of Universities for Research in Astronomy (AURA) under a cooperative agreement with the National Science Foundation, on behalf of the Gemini partnership of Argentina, Brazil, Canada, Chile, the Republic of Korea, and the United States of America. Based on observations obtained at the international Gemini Observatory (GN-2023B-Q-135, GN-2024A-Q-403), a program of NSF NOIRLab, which is managed by the Association of Universities for Research in Astronomy (AURA) under a cooperative agreement with the U.S. National Science Foundation on behalf of the Gemini Observatory partnership.
Time-domain research by the University of Arizona team and D.J.S.\ is supported by National Science Foundation (NSF) grants 2108032, 2308181, 2407566, and 2432036 and the Heising-Simons Foundation under grant \#2020-1864. S.V.\ and the UC Davis time-domain research team acknowledge support by NSF grants AST-2008108.

This work makes use of observations from the Las Cumbres Observatory network. The LCO team is supported by NSF grants AST-1911225 and AST-1911151. We acknowledge the support of the staff of Xinglong 2.16-m telescope. The work of X.W.is supported by the National Science Foundation of China (NSFC grants 12288102, 12033003), the Tencent Xplorer prize.  

This work was enabled by observations made from the Gemini North telescope, located within the Maunakea Science Reserve and adjacent to the summit of Maunakea. We are grateful for the privilege of observing the Universe from a place that is unique in both its astronomical quality and its cultural significance.

\facilities{ADS, LCOGT (SBIG, Sinistro), Sleaford:PROMPT, NED, Swift (UVOT), Las Cumbres Observatory (Sinistro, FLOYDS), Gemini:Gillett(GMOS), LBT(PEPSI), Keck I (LRIS), Lick Shane (Kast), WISeREP}

\software{{\tt DRAGONS} \citep{chris_simpson_2024_13274573}, astropy \citep{2013A&A...558A..33A,2018AJ....156..123A}, CMFGEN \citep{HillierMiller1998,HillierDessart2012}, FLOYDS pipeline \citep{Valenti14}, HEA-Soft \citep{HEA-Soft2014}, \texttt{lcogtsnpipe} \citep{Valenti16}, MatPLOTLIB \citep{mpl},WISeREP \citep{wiserep}, Spextool \citep{Cushing04}, LPipe \citep{2019PASP..131h4503P}}

\appendix

Table~\ref{tab:optspec} shows the complete optical spectroscopy log, and Table~\ref{tab:nirspec} shows the complete NIR spectroscopy log. All data will be made available as data behind the figure and are available on WISeRep (\url{https://www.wiserep.org})

 \begin{deluxetable*}{lcccccc}
\tablecaption{Optical Spectroscopy of SN~2024bch \label{tab:optspec}}
\tablehead{ \colhead{UT Date}    &\colhead{MJD}& \colhead{Phase}    &\colhead{Telescope+}   & \colhead{R}&  \colhead{Exposure Time}  \\[-6pt]
   \colhead{(y-m-d)}    &\colhead{} & \colhead{(days)} & \colhead{Instrument}  &\colhead{$\lambda$/$\Delta\lambda$}   & \colhead{(s)}   \  }
\startdata
2024-01-31 & 60340.45 & 2.8  &  Gemini-N+GMOS & 5800  & 600 $\times$ 4 \\
2024-01-31 & 60340.46 & 2.8 &  FTN+FLOYDS & 380 & 1800 \\
2024-02-01 & 60341.48 & 3.8 &  Gemini-N+GMOS & 930 & 300 $\times$ 4\\
2024-02-03 & 60343.54 & 5.9 &  FTN+FLOYDS & 380 & 900 \\
2024-02-04 & 60344.48 & 6.8 &  FTN+FLOYDS & 380 & 900 \\
2024-02-04 & 60344.62 & 7.0 &  Gemini-N+GMOS & 930 & 300 $\times$ 4\\
2024-02-05 & 60345.46 & 7.8  &  FTN+FLOYDS & 380 & 900 \\
2024-02-05 & 60345.52 & 7.9 &  LBT+PEPSI & 50000 &  5400\\
2024-02-06 & 60346.32 & 8.7 &  FTN+FLOYDS & 380 & 600 \\
2024-02-09 & 60349.57 & 11.9 &  FTN+FLOYDS & 380 & 600 \\
2024-02-11 & 60351.30 & 13.7 &  FTN+FLOYDS & 380 & 600 \\
2024-02-13 & 60353.30 & 15.7 &  Bok+B\&C & 700 & 600  $\times$ 3\\
2024-02-14 & 60354.37 & 16.7 &  FTN+FLOYDS & 380 & 600 \\
2024-02-16 & 60356.65 & 19.0 &  Xinglong+BFOSC & 400 & 2100 \\
2024-02-19 & 60359.34 & 21.7 &  FTN+FLOYDS & 380 & 600 \\
2024-02-19 & 60359.43 & 21.8 &  Gemini-N+GMOS & 930 & 300 $\times$ 4 \\
2024-02-22 & 60362.40 & 24.8 &  FTN+FLOYDS & 380 & 600 \\
2024-02-25 & 60365.49 & 27.9 &  FTN+FLOYDS & 380 & 600 \\
2024-02-28 & 60368.35 & 30.7 &  FTN+FLOYDS & 380 & 600 \\
2024-02-29 & 60369.41 & 31.8 &  Gemini-N+GMOS & 930 & 300 $\times$ 4 \\
2024-03-02 & 60371.10 & 33.5 &  Bok+B\&C &  700 & 600  $\times$ 3 \\
2024-03-02 & 60371.27 & 33.6 &  FTN+FLOYDS & 380 & 600 \\
2024-03-11 & 60380.23 & 42.6 &  LBT+PEPSI & 50000 &  14400\\
2024-03-17 & 60386.51 & 48.9 &  FTN+FLOYDS & 380 & 600 \\
2024-03-20 & 60389.30 & 51.7 &  Lick+KAST & 800 &  1200\\
2024-03-24 & 60393.43 & 55.8 &  Gemini-N+GMOS & 930 & 300 $\times$ 4 \\
2024-03-25 & 60394.31 & 56.7 &  FTN+FLOYDS & 380 & 900 \\
2024-03-29 & 60398.47 & 60.8 &  Xinglong+BFOSC & 400 & 3000 \\
2024-03-31 & 60400.36 & 62.7 &  FTN+FLOYDS & 380 & 900 \\
2024-04-03 & 60403.36 & 65.7 &  Gemini-N+GMOS & 930 & 300 $\times$ 4 \\
2024-04-06 & 60406.36 & 68.7 &  Keck+LRIS & 630/1700 & 300 \\
2024-04-08 & 60408.28 & 70.6 &  FTN+FLOYDS & 380 & 900 \\
2024-04-11 & 60411.27 & 73.6 &  Gemini-N+GMOS & 930 & 300 $\times$ 4 \\
2024-04-13 & 60413.10 & 75.5 &  Bok+B\&C & 700 & 900  $\times$ 3 \\
2024-04-15 & 60415.50 & 77.9 &  Xinglong+BFOSC & 400 & 3300 \\
2024-04-17 & 60417.24 & 79.6 &  FTN+FLOYDS & 380 & 900 \\
2024-04-19 & 60419.26 & 81.6 &  Lick+KAST & 800 & 2700 \\
2024-04-24 & 60424.39 & 86.7 &  FTN+FLOYDS & 380 & 900 \\
2024-04-30 & 60430.24 & 92.6 &  Gemini-N+GMOS & 930 & 300 $\times$ 4 \\
2024-05-02 & 60432.41 & 94.8 &  FTN+FLOYDS & 380 & 1800 \\
2024-05-10 & 60440.20 & 103 &  Bok+B\&C & 700 & 1500  $\times$ 3 \\
2024-05-24 & 60454.35 & 117 &  FTN+FLOYDS & 380 & 2700 \\
2024-06-01 & 60462.26 & 125 &  Gemini-N+GMOS & 930 & 300  $\times$ 2\\
2024-06-01 & 60462.33 & 125 &  FTN+FLOYDS & 380 & 2700 \\
2024-06-06 & 60467.27 & 130 &  FTN+FLOYDS & 380 & 2700 \\
2024-06-20 & 60481.26 & 144 &  Gemini-N+GMOS & 930 & 300  $\times$ 2\\
2024-06-23 & 60484.26 & 147  &  FTN+FLOYDS & 380 & 2700 \\
\hline
\enddata
 \tablecomments{Phases are with respect to an estimated explosion epoch of MJD 60337.65.}
 \end{deluxetable*}

 \begin{deluxetable*}{lccccc}
\tablecaption{NIR Spectroscopy of SN~2024bch \label{tab:nirspec}}
\tablehead{ \colhead{UT Date}    &\colhead{MJD}& \colhead{Phase}    &\colhead{Telescope+}   &  \colhead{Exposure Time}  \\[-6pt]
   \colhead{(y-m-d)}    &\colhead{} & \colhead{(days)} & \colhead{Instrument}     &    \  }
\startdata
2024-01-31 & 60340.24 & 2.6  &  MMT+MMIRS  &  4x180s \\
2024-02-20 & 60361.48 & 24  &  IRTF+SPeX  & 8x200s \\
2024-02-28 & 60369.40 & 32  &  IRTF+SPeX  & 8x200s \\
2024-03-28 & 60397.14 & 60  &  MMT+MMIRS  &  8x120s \\
\hline
\enddata
 \tablecomments{Phases are with respect to an estimated explosion epoch of MJD 60337.65.}
 \end{deluxetable*}

\bibliography{SN2024bch}
\end{document}